\documentclass[aps,prl,final,twocolumn,groupedaddress]{revtex4-1}

\pdfoutput=1

\usepackage{amsmath}
\usepackage{braket}
\usepackage{graphicx}
\usepackage{epsfig}
\usepackage{natbib}
\usepackage{color}

\newcommand{\iu}{\mathrm{i}}

\newcommand{\hn}{\hat{n}}
\newcommand{\hx}{\hat{x}}
\newcommand{\hy}{\hat{y}}
\newcommand{\hz}{\hat{z}}

\newcommand{\hno}{\hat{n}_0}

\newcommand{\he}{\hat{e}}

\newcommand{\Oi}{\Omega_1}
\newcommand{\Oii}{\Omega_2}
\newcommand{\Oio}{\Omega_{10}}
\newcommand{\Oiio}{\Omega_{20}}
\newcommand{\Oik}{\Omega_{1k_x}}
\newcommand{\Oiik}{\Omega_{2k_x}}

\newcommand{\ver}{\vec{r}}

\newcommand{\txi}{\tilde{\xi}}

\begin{document}


\title{Magnonic Goos-H\"anchen effect induced by one dimensional solitons}


\author{Victor Laliena}
\email[]{laliena@unizar.es}
\affiliation{Aragon Nanoscience and Materials Institute 
  (CSIC -- University of Zaragoza) and
  Condensed Matter Physics Department, University of Zaragoza \\
  C/Pedro Cerbuna 12, Zaragoza, 50009, Spain}

\author{Javier Campo}
\email[]{javier.campo@csic.es}
\affiliation{Aragon Nanoscience and Materials Institute 
  (CSIC -- University of Zaragoza) and
  Condensed Matter Physics Department, University of Zaragoza \\
  C/Pedro Cerbuna 12, Zaragoza, 50009, Spain}


\date{November 16, 2020}

\begin{abstract}
  The magnon spectral problem is solved in terms of the spectrum of a diagonalizable
  operator for a generic class of magnetic states that includes several types of domain walls
  and the chiral solitons of monoaxial helimagnets. Focusing on the isolated solitons of
  monoaxial helimagnets, it is shown that the spin waves scattered (reflected and transmitted)
  by the soliton suffer a lateral displacement analogous to the Goos-H\"anchen effect of optics.
  The displacement is a fraction of the wavelength, but can be greatly enhanced by using
  an array of well separated solitons.
  Contrarily to the Goos-H\"anchen effect recently studied in some magnetic systems, which takes
  place at interfaces between different magnetic systems, the effect
  predicted here takes place at the soliton position, what it is interesting  from the point
  of view of applications since solitons can be created at different places and moved across
  the material.
  This kind of Goos-H\"anchen effect is not particular of monoaxial helimagnets,
  but it is generic of a class of magnetic states, including domain walls in
  systems with interfacial Dzyaloshinskii-Moriya interaction.
\end{abstract}

\pacs{111222-k}

\maketitle


Magnonics is a subject of much interest in recent years since it is a promising
field that could transform the design of devices for information technology
\cite{Chumak15}.
Replacing electric currents by spin waves as information carriers in electronic devices
would imply a large reduction of heat production and energy consumption due to
the absence of Joule heating. Conceptual designs of devices based on spin waves
have already been proposed \cite{Chumak14,Schneider08}.
One of the main challenges with spin waves is its control
and manipulation. This control can be achieved in part by using the magnetic modulations
of nanometric scale that are (meta)stable in some materials: domain walls,
skyrmions, or chiral solitons. These solitonic states appear easily 
in chiral magnets, which are characterized by the presence of an important
Dzyaloshinskii-Moriya interaction (DMI).
Domain walls and their magnonics, with and without DMI, are being
extensively studied \cite{Winter61,Thiele73,Hertel04,LeMaho09,Garcia14,Kim16,
  Borys16,Whitehead17,Zingsem19}.
Comparatively, monoaxial helimagnets, in which the DMI acts only along one axis,
called the DMI axis, have received much less attention
\cite{Togawa12,Laliena16a,Laliena16b,Shinozaki16,Tsuruta16,Kishine16,Laliena17a,
  Goncalves17,Laliena18c,Masaki18,Kishine20,Laliena20}.

Generically, the magnonics of the non-collinear states faces
some mathematical difficulties related to the nature of the magnon wave equation.
The problem is not merely technical, but it raises the question of whether
a spectral representation for the spin waves exists in general, that is, whether
a general solution of the linearized Landau-Lisftchitz-Gilbert (LLG) equation can
be expressed as a combination of well defined spin wave modes.

In this work we develop a generic method that provides rigorously a complete
solution of the spectral magnon problem in terms of the spectrum of a
diagonalizable operator, for especial cases including the domain walls of
many systems and the isolated
soliton (IS) and the chiral soliton lattice (CSL) of monoaxial helimagnets.
As a by-product, by applying this formalism, we predict the existence of a
Goos-H\"anchen effect in the scattering of magnons by certain localized
one-dimensional magnetic modulated structures, such as solitons.
Before presenting this method we analyze a general problem
of magnonics, proving that the spectral representation of
spin waves does exist in general.

Consider a generic magnetic system described by a magnetization vector
field $\vec{M}=M_0\hn$,
with constant modulus, $M_0$, and direction given by the unit vector $\hn$.
Its energy is given by an energy functional $E[\hn]$.
The stationary states are those at which the variational derivative of $E[\hn]$
vanishes. The (meta)stable states are the local minima of $E[\hn]$,
a subset of the stationary states. Let $\hno$ be one stationary point of the energy.
Small fluctuations around $\hno$ can be written in terms of two real
fields, $\xi_1$ and $\xi_2$, writing 
\mbox{$\hn = (1-\xi^2)^{1/2}\hno + \xi_1\he_1 + \xi_2\he_2$},
where $\{\he_1,\he_2,\hno\}$ form an orthonormal triad.
These two fields are grouped into a two component field, a ``spinor'' $\txi$,
represented by the column matrix $\txi= [\xi_1, \xi_2]^\mathrm{T}$.
``Spinors'' are  denoted in this work by tilded
symbols \footnote{To avoid any misinterpretation, let us
  clarify that the term ``spinor'' is used here to distinguish the two-component
  object $\txi$ from one-component fields and three-dimensional vectors.
  Obviously, it is
  an abuse of language, since spinors are related to spatial rotations in a
  very precise way, very different from our $\txi$.}.
We use the notation \mbox{$(f,g) = \int d^3r \,f^*(\ver)\,g(\ver)$}
for the scalar product of two functions and
\mbox{$\langle\txi,\tilde{\eta}\rangle = (\xi_1,\eta_1) + (\xi_2,\eta_2)$}
for the scalar product of two ``spinors''.

Let us expand $E[\hn]$ in powers of $\xi_i$ to quadratic order:
\mbox{$E = E_0 + 2A(1/2)\langle\txi,K\txi\rangle + O(\xi^3)$}. 
The linear term vanishes since $\hno$ is a stationary state.
The constant $A$ has dimensions of energy per unit length and 
$K$ is a $2\times 2$ hermitian operator given
\begin{equation}
  K=\begin{pmatrix} K_{11} & K_{12} \\ K_{12}^\dagger & K_{22} \end{pmatrix},
\label{eq:K}
\end{equation}
where $K_{11}$ and $K_{22}$ are hermitian.
The $K_{ij}$ are integro-differential real operators.
If $\hno$ is (meta)stable, $K$ is positive (semi)definite.
This requires that both $K_{11}$ and $K_{22}$
be positive (semi)definite, and imposes constraints on $K_{12}$ that
we do not analyze here.

The oscillations of the magnetization about the (meta)stable state obey
the LLG equation, 
\mbox{$\partial_t\hn = \gamma\vec{B}_{\mathrm{eff}}\times\hn
  + \alpha\hn\times\partial_t\hn$},
where $\gamma$ is the gyromagnetic constant,
\mbox{$\vec{B}_{\mathrm{eff}}=-\delta E/\delta\hn$} is the effective field,
and $\alpha$ the Gilbert damping parameter.
We ignore the damping and set $\alpha=0$ in the remaining of the paper.
Let us pick up some characteristic parameter of the system with units of
inverse length, $q_0$, and introduce the constant
$\omega_0=\gamma 2Aq_0^2/M_0$, with dimensions of inverse time.
Considering small oscillations, we expand the LLG equation in powers
of $\xi$ around $\hno$.
The zero-th order term vanishes since $\hno$ is a stationary point.
To linear order we obtain 
\mbox{$\partial_t\txi=\Omega\txi$}, where \mbox{$\Omega=(\omega_0/q_0^2)JK$},
with 
\begin{equation}
  J=\begin{pmatrix} 0 & -I \\ I & 0 \end{pmatrix}.
\label{eq:J}
\end{equation}
In the above expression $I$ is the identity operator.

$\Omega$ is not anti-hermitian (not even normal), what
raises the issues mentioned before about the spectral properties
of the spin waves, like the existence of a complete set of well defined
modes with definite frequency. We provide here a general formal answer.
The spectral equation is
\mbox{$\Omega\txi=\nu\txi$}, with $\nu$ a complex
eigenvalue. For a (meta)stable state the square root of $K$ is a well defined
hermitian positive (semi)definite operator.
Multiplying both sides of the spectral equation by
$K^{1/2}\Omega$ we obtain
\begin{equation}
(\omega_0^2/q_0^4) \big(K^{1/2}JKJK^{1/2}\big)\,K^{1/2}\,\txi=\nu^2\,K^{1/2}\txi.
\end{equation}
Hence, the spectral properties of $\Omega$ are derived from the spectral
properties of \mbox{$K^{1/2}JKJK^{1/2}$}, which is hermitian, and therefore has
a complete set of orthogonal eigenstates, denoted by $\{\tilde{\eta}_i\}$. Then
$\{\txi_i=K^{-1/2}\tilde{\eta}_i\}$ is a complete set of eigenstates
of $\Omega$, which satisfy the normalization condition
\mbox{$\langle\txi_i,K\txi_j\rangle = \delta_{ij}$}.
It is easily checked that \mbox{$K^{1/2}JKJK^{1/2}$} is negative (semi)definite,
so that $\nu^2\leq0$, and $\nu=\iu\omega$, with $\omega$ real.
Thus, for a (meta)stable state, the spectrum of $\Omega$ lies on the
imaginary axis and its eigenstates form a complete set
\footnote{If $K$ has zero modes, $K^{-1/2}$ is not defined, and this
  argument is problematic, but it could be modified to circumvent the problem}.

The spectral problem for $\Omega$ is easy if the four operators
$K_{ij}$ commute, as in ferromagnetic (FM), helical, and conical
states \cite{Laliena17b},
and in some domain walls \cite{Winter61}.
In those cases the problem is reduced to find the spectrum of
one hermitian operator ($K_{11}$ for instance) and the diagonalization
of a $2\times 2$ matrix.

In what follows, we address problems in which the $K_{ij}$ do not commute,
focusing on the cases were $K_{12}=0$, for which we give a complete
solution. Examples include the IS and the CSL
of monoaxial helimagnets, and the domain walls of some systems
with DMI \cite{Borys16}.
In this last instance the authors addressed the problem via perturbation theory,
splitting $\Oii$ as the sum of an operator that commutes with $\Oi$ plus a
perturbation.
This may be a reasonably approach, especially if the unperturbed operator
can be treated analytically, provided it can be guaranteed that the
perturbation does not originate new bound states.

Let us define \mbox{$\Oi=(\omega_0/q_0^2)K_{11}$} and
\mbox{$\Oii=(\omega_0/q_0^2)K_{22}$}.
As shown above, the eigenvalues of $\Omega$ for a (meta)stable state
are purely imaginary, $\iu\omega$, with $\omega$ real.
In components, the spectral equation for $\Omega$ gives
\mbox{$\Oii\,\xi_2 = -\iu\omega\,\xi_1$} and
\mbox{$\Oi\,\xi_1 = \iu\omega\,\xi_2$}.
Substituting the values of $\xi_1$ and $\xi_2$ given explicitly by one of
these equations into the other, we obtain
\mbox{$\Oii\Oi\xi_1 = \omega^2 \xi_1$} and \mbox{$\Oi\Oii\xi_2 = \omega^2 \xi_2$}.
These two equations are compatible since $\Oii\Oi$ and $\Oi\Oii$ have
the same spectrum: if $\xi_1$ is an eigenfunction of $\Oii\Oi$
then $\Oi\xi_1$ is an eigenfunction of $\Oi\Oii$ with the same eigenvalue;
the same is true changing 1 by 2.
The case $\omega=0$ is special: if $\xi_1$ is an
eigenfunction of $\Oii\Oi$ with zero eigenvalue, we have an eigenstate of $\Omega$ 
just taking $\xi_2=0$. Again, the statement is valid changing 1 by 2.

The $K$ operator of a (metas)stable state may be gapless or even have zero modes.
When $K_{12}=0$ the zero modes or the gapless modes are generically associated to
one operator, say $K_{11}$, and $K_{22}$ has a gap.
Hence $\Oii$ is a hermitian positive definite invertible operator, and so it is
its square root. Therefore, although $\Oii\Oi$ is not hermitian (not even normal),
equation \mbox{$\Oii\Oi\xi_1 = \omega^2 \xi_1$} can be written
in terms of the hermitian positive semidefinite operator
\mbox{$\Lambda=\Oii^{1/2}\,\Oi\,\Oii^{1/2}$} as
\mbox{$\Lambda\,\big(\Oii^{-1/2}\xi_1\big) = \omega^2\,\big(\Oii^{-1/2}\xi_1\big)$}.
Therefore, the spectral problem for $\Omega$ is completely solved in terms of
the spectral problem 
\mbox{$\Lambda\upsilon = \omega^2\upsilon$},
just setting $\xi_1=\Oii^{1/2}\upsilon$ and $\xi_2=\Oii^{-1/2}\upsilon$,
where we used the equation \mbox{$\Oii\,\xi_2 = -\iu\omega\,\xi_1$}.
If $\{\upsilon_i\}$ is a complete set of orthonormal eigenfunctions of $\Lambda$,
then \mbox{$\{\psi_i=\omega_0^{-1/2}\,\Oii^{1/2}\,\upsilon_i$\}} is
a complete set of eigenfunctions of $\Oii\,\Oi$ that satisfy the condition
\begin{equation}
(\psi_i,\,\omega_0\,\Oii^{-1}\,\psi_j) = N_i\,\delta_{ij}, \label{eq:norm1}
\end{equation}
where $N_i$ provides a proper normalization condition \footnote{These results
  can be easily obtained by noticing that $\Oii\Oi$ is a hermitian
  positive (semi)definite operator with respect to the scalar product 
  \mbox{$((f,g)) = (f,\omega_0\Oii^{-1}g)$}.
  Therefore, the eigenvalues of $\Oii\Oi$ are real and non-negative
  and its eigenfunctions are ortohogonal with respect to the $((,))$
  product, what amounts to Eq.~(\ref{eq:norm1}).}.

We find it convenient to express the eigenstates of $\Omega$ in terms of
the eigenfunctions of $\Oii\Oi$,  $\psi_i$. Since $\Omega$ is real, its spectrum comes in
complex conjugate pairs. Hence,
each $\psi_i$ gives rise to two eigenstates of $\Omega$,
with eigenvalues $\iu\sigma\omega_i$, with $\omega_i\geq 0$ and $\sigma=\pm 1$,
given by
\begin{equation}
  \txi^{(i\,\sigma)} = \frac{1}{(1+\omega_i^2/\omega_0^2)^{1/2}}
  \begin{pmatrix} \psi_i \\ -\iu\sigma\omega_i\Oii^{-1}\psi_i \end{pmatrix},
  \label{eq:evec}
\end{equation} 
which satisfy the normalization condition
\begin{equation}
  \langle\txi^{(i\,\sigma)},G\txi^{(j\,\sigma^\prime)}\rangle =
  \frac{\omega_0^2+\sigma\sigma^\prime\omega_i^2}{\omega_0^2+\omega_i^2}N_i\delta_{ij},
  \label{eq:norm2}
\end{equation}
where
\begin{equation}
G=\begin{pmatrix} \omega_0\Oii^{-1} & 0 \\ 0 & \omega_0^{-1}\Oii \end{pmatrix}.
\label{eq:G}
\end{equation}
The completitude of the set $\{\psi_i\}$ implies the completitude of the set
$\{\txi^{(i\sigma)}\}$:
for any given $\txi$ we have \mbox{$\txi = \sum_{i\sigma} c_{i\sigma} \txi^{(i\sigma)}$},
where, defining $\bar{\sigma}=-\sigma$,
\begin{equation}
  \hspace*{-0.2cm}
  c_{i\sigma} \!= \!\frac{\big(\omega_i^2+\omega_0^2\big)^2}{4N_i\omega_0^2\omega_i^2}
    \!\left[\!
       \big\langle\txi^{(i\sigma)},G\txi\big\rangle \!
       + \!\frac{\omega_i^2-\omega_0^2}{\omega_i^2+\omega_0^2}\!
       \big\langle\txi^{(i,\bar{\sigma})},G\txi\big\rangle\!
    \right]\!. \hspace*{-0.2cm}
  \label{eq:cij}
\end{equation}

In summary, we have obtained the eigenstates $\txi^{i\sigma}$ of $\Omega$ in terms of the
eigenfunctions $\psi_i$ of the diagonalizable operator $\Oii\Oi$, for the cases in which
$K_{12}=0$, what allows to solve a number of important problems.
Moreover, Eqs. (\ref{eq:evec})-(\ref{eq:cij}) can be taken as a starting point to quantization,
by imposing canonical commutation relations to $\xi_1$ and $\xi_2$, which are derived
from the algebra of angular momentum satisfied by the quantized components of $\hn$.

In the following we apply this method to the case of an IS in a monoaxial
helimagnet, which is characterized by an energy functional
\mbox{$E[\hn] = 2A\int d^3r W$}, with
\begin{equation}
W = \frac{1}{2} \partial_i\hn\cdot\partial_i\hn 
-q_0\hz\cdot(\hn\times\partial_z\hn)
- \frac{1}{2}q_0^2\kappa(\hz\cdot\hn)^2
- q_0^2\vec{h}\cdot\hn. \label{eq:W}
\end{equation}
The successive terms of the right-hand side represent a FM exchange interaction, a uniaxial DMI
along the $\hz$ axis, an easy-plane ($\kappa<0$) uniaxial magnetic anisotropy (UMA) along the DMI
axis, and a Zeeman interaction with an external magnetic field perpendicular to the
DMI axis, with $\vec{h}=h\hy$.
For simplicity, we ignore the magnetostatic energy.
The constant $q_0$ is proportional to the ratio between the DMI and FM exchange interaction
strengths, and plays the role of the $q_0$ parameter
introduced above, and $\kappa$ and $\vec{h}$ are dimensionless.
The numerical results discussed below correspond to $\kappa=-5.0$ and $h=1.0$, unless other values are
explicitly quoted.

The Sine-Gordon soliton is a stationary point, given by
\mbox{$\hno=-\sin\varphi\,\hx+\cos\varphi\,\hy$}, with
\mbox{$\varphi(z) = 4 \arctan[\exp(z/\Delta)]$},
where \mbox{$\Delta=1/(q_0\sqrt{h})$} is the soliton width.
Notice that $\hno$ is confined to the plane perpendicular to the DMI axis.
The solitons are metastable below a certain value of $h$ that depends on the
DMI and UMA strengths \cite{Laliena20}, and they condense into a CSL for $h$ below the
critical field $h_c=\pi^2/16$ \cite{Dzyal64}. 

Taking $\he_1=\hz\times\hno$ and $\he_2=\hz$, so that $\xi_1$ and $\xi_2$ describe 
the in-plane and out-of-plane oscillations, respectively, the operators $\Oi$ and $\Oii$ are given by
\begin{eqnarray}
\Oi &=& \frac{\omega_0}{q_0^2}\big[-\nabla^2 + U_1 + q_0^2h\big], \\
\Oii &=& \frac{\omega_0}{q_0^2}\big[-\nabla^2 + U_2 + q_0^2(h-\kappa)\big], 
\end{eqnarray}
where \mbox{$U_1= -(1/2)\varphi^{\prime\,2}$} and
\mbox{$U_2 = -(3/2)\varphi^{\prime\,2}+2q_0\varphi^\prime$}
are even functions of $z$ and decay exponentially to zero
when $z\rightarrow\pm\infty$, since \mbox{$\varphi^\prime(z)=2/[\Delta\cosh(z/\Delta)]$}.
These functions are independent of $\kappa$, but depend on $h$ through $\Delta$.

The operators $\Oi$ and $\Oii$ are partially diagonalized by a Fourier
transform in $x$ and $y$. Since $x$ and $y$ enter the problem in a symmetric
way, to simplify the notation we consider only the $x$ dependence,  
writing the eigenfunctions of $\Oii\Oi$ as
\mbox{$\psi_{k_x}(x,z) = \exp(\iu k_x x) \phi_{k_x}(z)$}.
The general case is obtained by replacing $k_x^2$ by $k_x^2+k_y^2$ and
$\exp(\iu k_x x)$ by \mbox{$\exp[\iu (k_x x+k_y y)]$}.
After the Fourier transform, the spectral problem becomes
\mbox{$\Oiik\Oik\phi_{k_x}=\omega^2\phi_{k_x}$},
where $\omega^2$ is a function of $k_x^2$ and
\begin{equation}
  \Oik = \Oio + \omega_0 k_x^2/q_0^2, \quad \Oiik = \Oiio + \omega_0 k_x^2/q_0^2,
\end{equation}
with $\Oio$ and $\Oiio$ obtained by replacing $\nabla^2$ by $\partial_z^2$ in
$\Oi$ and $\Oii$.
The eigenfunctions, $\phi_{k_xi}$, labeled by $i$,
satisfy a normalization condition analogous to (\ref{eq:norm1}).

Non-reciprocal propagation, usually associated to chirality, is absent in the IS and in the CSL,
because it require first order derivatives in the $\Omega$ operator, which is not the case.
It is easy to see, by deriving the generic form of the $K$ operator associated to (\ref{eq:W}),
that non-reciprocal propagation takes place in monoaxial helimagnets only
in states whose magnetic moments have a non-vanishing projection onto the DMI axis.

\begin{figure}[t!]
  \centering
  \hspace*{-0.6cm}
  \includegraphics[width=0.255\textwidth]{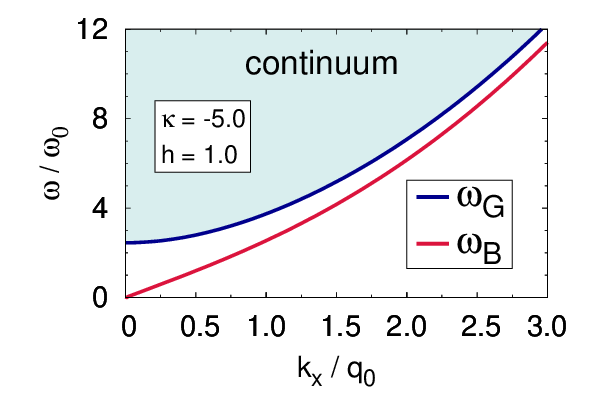}\* \hspace{-0.6cm}
  \includegraphics[width=0.255\textwidth]{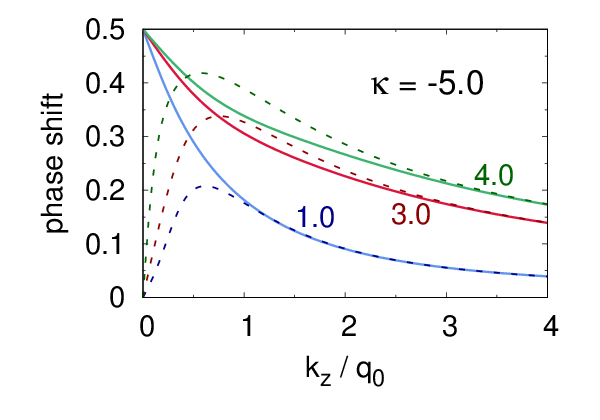}\*
  \vspace*{-0.4cm}
  \caption{
    Left: spin wave spectrum. The red line is the dispersion relation,
    $\omega_\mathrm{B}(k_x)$, for the bound state branch,
    and the blue line signals the gap.
    Right: Phase shifts $\delta_0$ (continuous lines) and $\delta_1$ (broken lines)
    \textit{vs.} $k_z$ for $k_x=0$ for the indicated values of $h$.
    \label{fig:fig1}}
\end{figure}

The spectral problems were solved numerically for a large discrete set of $k_x$,
on a box $-L\leq z \leq L$ with Dirichlet boundary conditions at $z=\pm L$ \cite{suppl}.
Insight about the spectrum is obtained
by studying the asymptotic properties of the eigenfunctions as $z\rightarrow\pm\infty$,
given in the supplemental material \cite{suppl}. 
The spectrum, depicted in Fig.~\ref{fig:fig1} (right),
contains a continuum of states unbounded in all directions,
with frequencies above a gap given by
\begin{equation}
  \omega_\mathrm{G}(k_x) = \omega_0\big[(k_x^2/q_0^2+h)(k_x^2/q_0^2+h-\kappa)\big]^{1/2},
\end{equation}
which is obtained by standard means from the asymptotic analysis \cite{suppl}.
Below the gap there is a gapless branch of states,
consisting of waves bounded to the
soliton position, that is, decaying exponentially as $z\rightarrow\pm\infty$, but
unbounded in the other directions. 

We shall analyze the gapless branch elsewhere. Here we focus on the continuum states,
that are used to describe the scattering of a magnon wave packet by the soliton,
which results in the emergence of one reflected and one transmitted wave packet
(the scattered waves). Although $\Oiik\Oik$ is not hermitian,
nor second order in derivatives, the concepts of scattering theory
are valid since they rely only on the asymptotic properties
of the wave equation \cite{Galindo90I}.
This allows us to predict one unusual feature of the scattering: the Goos-H\"anchen effect.

The eigenfunctions $\Oiik\Oik$ are either even or odd functions
of $z$, due to the $z\rightarrow -z$ invariance.
The continuum states are degenerate,
and for each $\omega^2$ there is an even and an odd eigenfunction,
behaving as $z\rightarrow\pm\infty$ as 
\begin{equation}
  \phi^{(e)}_{k_xk_z}(z) \sim \cos(k_zz\pm\delta_0), \;\;
  \phi^{(o)}_{k_xk_z}(z) \sim \sin(k_zz\pm\delta_1), \label{eq:ass}
\end{equation}
where the superscripts $e$ and $o$ stand for even and odd, respectively,
and $\delta_0$ and $\delta_1$ are the corresponding phase shifts,
which depend on $k_x$ and $k_z$.
The wave number $k_z$ is obtained from $\omega^2$ using the dispersion relation \cite{suppl}:
\begin{equation}
  k_z = q_0\left[\left(\frac{\omega^2}{\omega_0^2}+\frac{\kappa^2}{4}\right)^{1/2}
    -\left(\frac{k_x^2}{q_0^2}+h-\frac{\kappa}{2}\right)\right].
\end{equation}

The phase shifts are obtained by combining the asymptotic
behaviour of Eqs. (\ref{eq:ass}) and the boundary condition at
$z=L$, what gives \mbox{$k_zL + \delta_i = 2\pi n_i$}, for $i=0,1$, 
where $n_i$ are integers and \mbox{$\delta_i\in[-\pi,\pi]$}.
The phase shifts for $k_x=0$ are shown as a function of $k_z$ in
Fig. \ref{fig:fig1} (right).
In contrast with the the domain wall case \cite{Winter61}, which is
reflectionless for magnons, the reflection coefficient,
\mbox{$R=\sin^2(\delta_0-\delta_1)$}, does not vanish since $\delta_0\neq\delta_1$.

It is curious that, in spite that it has been demonstrated only for some
classes of Schr\"odinger operators, and $\Oiik\Oik$ is not a
Schr\"odinger operator, the phase shifts agree with the thesis of Levinson
theorem \cite{Galindo90I,Levinson49}, which states that
\mbox{$[\delta_0(0)-\delta_0(\infty)]/\pi+1/2$}
and \mbox{$[\delta_1(0)-\delta_1(\infty)]/\pi$}
are equal to the number of bound states of the respective parities.
The agreement follows from $\delta_0=\pi/2$ and $\delta_1=0$ for $k_z=0$,
$\delta_0=\delta_1=0$ for $k_z\rightarrow\infty$, and
the existence of a single bound state (gapless branch), which is even.

\begin{figure}[t!]
  \centering
  \hspace{-0.5cm}
  \includegraphics[width=0.25\textwidth]{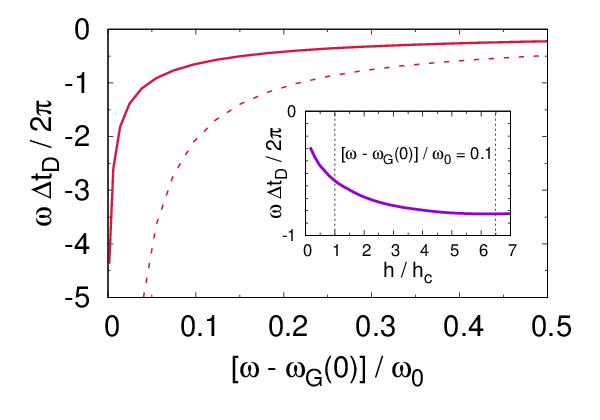}\* \hspace{-0.4cm}
  \includegraphics[width=0.25\textwidth]{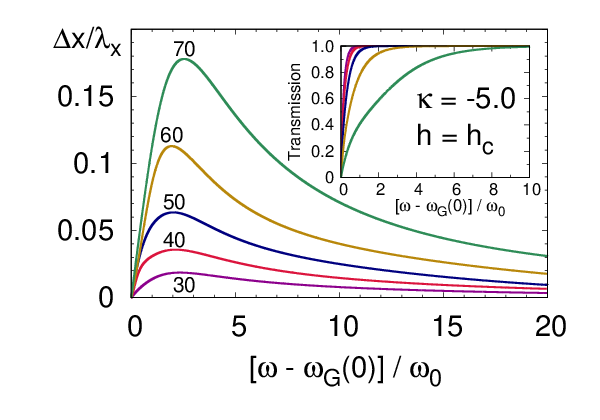}\*
  \vspace*{-0.25cm}
  \caption{Left: time delay in units of the wave period \textit{vs.} frequency
    \mbox{$\omega-\omega_\mathrm{G}(0)$}.
    The broken line is the Wigner causal bound.
    Inset: time delay \textit{vs.} $h/h_c$ for $\omega-\omega_\mathrm{G}(0)=0.1\,\omega_0$.
    Right: Goos-H\"anchen shift for several incidence angles, in degrees,
    for $h=h_c$.
    Inset: transmission coefficient for the same angles.
    \label{fig:fig2}} 
\end{figure}

The dependence of the phase shifts on the frequency introduces a time delay in the
scattered (reflected and transmitted) waves given by 
\mbox{$\Delta t_D = d(\delta_0+\delta_1)/d\omega$} \cite{Wigner55}.
It is indeed an advance time, since we obtain $\Delta t_D<0$. This is usually the
case when the scattering potential is repulsive, so that we may conclude that the soliton
repels the magnons. It was shown by Wigner that
causality implies the bound $\Delta t_D \geq - (2ak+1)/kv$, where $a$ is the range
of the potential,
$k^2=k_x^2+k_z^2$, and $v=d\omega/dk$ is the group
velocity \cite{Wigner55}.
In our case we may reasonably estimate the bound taking $a=\Delta$.
The product $\omega\Delta t_D$ \textit{versus} \mbox{$\omega-\omega_G$} is shown in
Fig.~\ref{fig:fig2} (left) for \mbox{$k_x=0$}.
The Wigner bound (broken line) is well satisfied.
The delay time is appreciable for frequencies close to $\omega_G$ and, as
the inset shows, decreases with the magnetic field strength.

The non trivial dependence of $\Oiik\Oik$ on $k_x$ induces a $k_x$ dependence of
the phase shifts, which originates a displacement of the scattered waves
(reflected and transmitted) perpendicular to $\hz$. That is, if the center of
a wave packet of narrow cross section impinges the soliton at a point $x$,
the scattered wave packets
left the soliton centered at a point $x+\Delta x$, where
\mbox{$\Delta x = -\partial (\delta_0+\delta_1)/\partial k_x$}.
This relation is derived from a stationary phase analysis of
the scattered wave \cite{Artmann48}.
This very interesting effect is analogous to the  well known Goos-H\"anchen 
effect of optics \cite{Goos47}, in which a light beam reflected
at the interface of two different media suffers a lateral
displacement given by an expression similar to the above $\Delta x$.
Recently, the Goos-H\"anchen effect for spin waves has been theoretically studied at
interfaces that separate different magnetic media
\cite{Dadoenkova12,Gruszecki14,Gruszecki15,Gruszecki17,Mailyan17,Wang19,Zhen20},
and experimental evidence of the effect at the edge of a Permalloy film has been
reported \cite{Stigloher18}.
To our knowledge, the kind of Goos-H\"anchen effect predicted here,
induced by a magnetic modulation instead of an interface, has not been considered before. 

The Goos-H\"anchen shift produced by magnetic modulations (not by interfaces)
is due to the non-commutativity of $\Oik$ and $\Oiik$: if they commute,
the phase shifts are independent of $k_x$, since then the eigenfunctions of
$\Oik\Oiik$ are the eigenfunctions of $\Oik$ or $\Oiik$, which
are independent of $k_x$, because $k_x$ enters this operators through a
multiple of the identity.
Examples in which $\Oik$ and $\Oiik$ do commute are the usual domain walls \cite{Winter61},
which therefore do not induce the Goos-H\"anchen effect.
The addition of an interfacial DMI, as in the model studied by
Borys \textit{et al.} \cite{Borys16}, spoils the commutativity of $\Oik$ and $\Oiik$
and therefore induce a Goos-H\"anchen effect in this kind of domain walls.
Borys \textit{et al.} did not address this question since they consider only the
propagation of spin waves in one dimension. To our knowledge, the Goos-H\"anchen effect
has not been analyzed yet in domain walls, in spite that it has to appear in some of them
(\textit{e.g.} those with DMI).
It can be done following the ideas presented in this work.

The shift $\Delta x$ that we obtain for the IS in a monoaxial
helimagnet is a fraction of the wavelength in the $\hx$ direction,
\mbox{$\lambda_x=2\pi/k_x$}.
This is very interesting because it opens the possibility of manipulating the spin waves
at the sub-wavelength scale.
Moreover, $\Delta x$ is additive as the wave is transmitted across an array of well
separated solitons, and therefore the shift can be enhanced by a large factor, provided
the transmission coefficient is high enough. The magnitude of the shift decreases with
the magnetic field, which acts as a control parameter.
Fig. \ref{fig:fig2} (right) displays $\Delta x/\lambda_x$ as a function of frequency
(relative to $\omega_\mathrm{G}$) for several values of the incidence angle, for the critical
field $h=h_c$. At this value of $h$ solitons can be easily created.
The inset shows the transmission coefficient for the same angles. We see that
there is a range of frequencies and incidence angles where $\Delta x/\lambda_x\approx 0.1$
and the transmission coefficient is very close to one, so that $\Delta x$ can be enhanced
to several tens of wavelengths.

To conclude, it is worthwhile to stress that the Goos-H\"anchen displacement predicted
here is not particular of monoaxial helimagnets, but it is expected in any one-dimensional
soliton for which $\Oio$ and $\Oiio$ do not commute, for instance in domain
walls with DMI \cite{Borys16}. It is also remarkable that it does not take place at the
interface between two different magnetic media, but at the soliton position.
For potential applications, this has the advantage that solitons can be created at
different locations and moved across the material by the application of magnetic fields
or polarized currents \cite{Laliena20}.





\begin{acknowledgments}
  Grants No PGC-2018-099024-B-I00-ChiMag from the Ministry of Science and Innovation of Spain,
  i-COOPB20524 from CSIC, DGA-M4 from the Diputaci\'on General de Arag\'on, are acknowledged.
\end{acknowledgments}


\bibliography{references}

\begin{thebibliography}{43}%
\makeatletter
\providecommand \@ifxundefined [1]{%
 \@ifx{#1\undefined}
}%
\providecommand \@ifnum [1]{%
 \ifnum #1\expandafter \@firstoftwo
 \else \expandafter \@secondoftwo
 \fi
}%
\providecommand \@ifx [1]{%
 \ifx #1\expandafter \@firstoftwo
 \else \expandafter \@secondoftwo
 \fi
}%
\providecommand \natexlab [1]{#1}%
\providecommand \enquote  [1]{``#1''}%
\providecommand \bibnamefont  [1]{#1}%
\providecommand \bibfnamefont [1]{#1}%
\providecommand \citenamefont [1]{#1}%
\providecommand \href@noop [0]{\@secondoftwo}%
\providecommand \href [0]{\begingroup \@sanitize@url \@href}%
\providecommand \@href[1]{\@@startlink{#1}\@@href}%
\providecommand \@@href[1]{\endgroup#1\@@endlink}%
\providecommand \@sanitize@url [0]{\catcode `\\12\catcode `\$12\catcode
  `\&12\catcode `\#12\catcode `\^12\catcode `\_12\catcode `\%12\relax}%
\providecommand \@@startlink[1]{}%
\providecommand \@@endlink[0]{}%
\providecommand \url  [0]{\begingroup\@sanitize@url \@url }%
\providecommand \@url [1]{\endgroup\@href {#1}{\urlprefix }}%
\providecommand \urlprefix  [0]{URL }%
\providecommand \Eprint [0]{\href }%
\providecommand \doibase [0]{http://dx.doi.org/}%
\providecommand \selectlanguage [0]{\@gobble}%
\providecommand \bibinfo  [0]{\@secondoftwo}%
\providecommand \bibfield  [0]{\@secondoftwo}%
\providecommand \translation [1]{[#1]}%
\providecommand \BibitemOpen [0]{}%
\providecommand \bibitemStop [0]{}%
\providecommand \bibitemNoStop [0]{.\EOS\space}%
\providecommand \EOS [0]{\spacefactor3000\relax}%
\providecommand \BibitemShut  [1]{\csname bibitem#1\endcsname}%
\let\auto@bib@innerbib\@empty
\bibitem [{\citenamefont {Chumak}\ \emph {et~al.}(2015)\citenamefont {Chumak},
  \citenamefont {Vasyuchka}, \citenamefont {Serga},\ and\ \citenamefont
  {Hillebrands}}]{Chumak15}%
  \BibitemOpen
  \bibfield  {author} {\bibinfo {author} {\bibfnamefont {A.~V.}\ \bibnamefont
  {Chumak}}, \bibinfo {author} {\bibfnamefont {V.~I.}\ \bibnamefont
  {Vasyuchka}}, \bibinfo {author} {\bibfnamefont {A.~A.}\ \bibnamefont
  {Serga}}, \ and\ \bibinfo {author} {\bibfnamefont {B.}~\bibnamefont
  {Hillebrands}},\ }\href {\doibase https://doi.org/10.1038/nphys3347}
  {\bibfield  {journal} {\bibinfo  {journal} {Nature Phys}\ }\textbf {\bibinfo
  {volume} {11}},\ \bibinfo {pages} {453} (\bibinfo {year} {2015})}\BibitemShut
  {NoStop}%
\bibitem [{\citenamefont {Chumak}\ \emph {et~al.}(2014)\citenamefont {Chumak},
  \citenamefont {Serga},\ and\ \citenamefont {Hillebrands}}]{Chumak14}%
  \BibitemOpen
  \bibfield  {author} {\bibinfo {author} {\bibfnamefont {A.~V.}\ \bibnamefont
  {Chumak}}, \bibinfo {author} {\bibfnamefont {A.~A.}\ \bibnamefont {Serga}}, \
  and\ \bibinfo {author} {\bibfnamefont {B.}~\bibnamefont {Hillebrands}},\
  }\href {\doibase https://doi.org/10.1038/ncomms5700} {\bibfield  {journal}
  {\bibinfo  {journal} {Nature Commun}\ }\textbf {\bibinfo {volume} {5}},\
  \bibinfo {pages} {4700} (\bibinfo {year} {2014})}\BibitemShut {NoStop}%
\bibitem [{\citenamefont {Schneider}\ \emph {et~al.}(2008)\citenamefont
  {Schneider}, \citenamefont {Serga}, \citenamefont {Leven}, \citenamefont
  {Hillebrands}, \citenamefont {Stamps},\ and\ \citenamefont
  {Kostylev}}]{Schneider08}%
  \BibitemOpen
  \bibfield  {author} {\bibinfo {author} {\bibfnamefont {T.}~\bibnamefont
  {Schneider}}, \bibinfo {author} {\bibfnamefont {A.~A.}\ \bibnamefont
  {Serga}}, \bibinfo {author} {\bibfnamefont {B.}~\bibnamefont {Leven}},
  \bibinfo {author} {\bibfnamefont {B.}~\bibnamefont {Hillebrands}}, \bibinfo
  {author} {\bibfnamefont {R.~L.}\ \bibnamefont {Stamps}}, \ and\ \bibinfo
  {author} {\bibfnamefont {M.~P.}\ \bibnamefont {Kostylev}},\ }\href {\doibase
  https://doi.org/10.1063/1.2834714} {\bibfield  {journal} {\bibinfo  {journal}
  {Appl. Phys. Lett.}\ }\textbf {\bibinfo {volume} {92}},\ \bibinfo {pages}
  {022505} (\bibinfo {year} {2008})}\BibitemShut {NoStop}%
\bibitem [{\citenamefont {Winter}(1961)}]{Winter61}%
  \BibitemOpen
  \bibfield  {author} {\bibinfo {author} {\bibfnamefont {J.~M.}\ \bibnamefont
  {Winter}},\ }\href@noop {} {\bibfield  {journal} {\bibinfo  {journal} {Phys.
  Rev.}\ }\textbf {\bibinfo {volume} {124}},\ \bibinfo {pages} {452} (\bibinfo
  {year} {1961})}\BibitemShut {NoStop}%
\bibitem [{\citenamefont {Thiele}(1973)}]{Thiele73}%
  \BibitemOpen
  \bibfield  {author} {\bibinfo {author} {\bibfnamefont {A.~A.}\ \bibnamefont
  {Thiele}},\ }\href {\doibase 10.1103/PhysRevB.7.391} {\bibfield  {journal}
  {\bibinfo  {journal} {Phys. Rev. B}\ }\textbf {\bibinfo {volume} {7}},\
  \bibinfo {pages} {391} (\bibinfo {year} {1973})}\BibitemShut {NoStop}%
\bibitem [{\citenamefont {Hertel}\ \emph {et~al.}(2004)\citenamefont {Hertel},
  \citenamefont {Wulfhekel},\ and\ \citenamefont {Kirschner}}]{Hertel04}%
  \BibitemOpen
  \bibfield  {author} {\bibinfo {author} {\bibfnamefont {R.}~\bibnamefont
  {Hertel}}, \bibinfo {author} {\bibfnamefont {W.}~\bibnamefont {Wulfhekel}}, \
  and\ \bibinfo {author} {\bibfnamefont {J.}~\bibnamefont {Kirschner}},\ }\href
  {\doibase 10.1103/PhysRevLett.93.257202} {\bibfield  {journal} {\bibinfo
  {journal} {Phys. Rev. Lett.}\ }\textbf {\bibinfo {volume} {93}},\ \bibinfo
  {pages} {257202} (\bibinfo {year} {2004})}\BibitemShut {NoStop}%
\bibitem [{\citenamefont {Le~Maho}\ \emph {et~al.}(2009)\citenamefont
  {Le~Maho}, \citenamefont {Kim},\ and\ \citenamefont {Tatara}}]{LeMaho09}%
  \BibitemOpen
  \bibfield  {author} {\bibinfo {author} {\bibfnamefont {Y.}~\bibnamefont
  {Le~Maho}}, \bibinfo {author} {\bibfnamefont {J.-V.}\ \bibnamefont {Kim}}, \
  and\ \bibinfo {author} {\bibfnamefont {G.}~\bibnamefont {Tatara}},\ }\href
  {\doibase 10.1103/PhysRevB.79.174404} {\bibfield  {journal} {\bibinfo
  {journal} {Phys. Rev. B}\ }\textbf {\bibinfo {volume} {79}},\ \bibinfo
  {pages} {174404} (\bibinfo {year} {2009})}\BibitemShut {NoStop}%
\bibitem [{\citenamefont {Garcia-Sanchez}\ \emph {et~al.}(2016)\citenamefont
  {Garcia-Sanchez}, \citenamefont {Borys}, \citenamefont {Vansteenkiste},
  \citenamefont {Kim},\ and\ \citenamefont {Stamps}}]{Garcia14}%
  \BibitemOpen
  \bibfield  {author} {\bibinfo {author} {\bibfnamefont {F.}~\bibnamefont
  {Garcia-Sanchez}}, \bibinfo {author} {\bibfnamefont {P.}~\bibnamefont
  {Borys}}, \bibinfo {author} {\bibfnamefont {A.}~\bibnamefont
  {Vansteenkiste}}, \bibinfo {author} {\bibfnamefont {J.-V.}\ \bibnamefont
  {Kim}}, \ and\ \bibinfo {author} {\bibfnamefont {R.~L.}\ \bibnamefont
  {Stamps}},\ }\href@noop {} {\bibfield  {journal} {\bibinfo  {journal} {Phys.
  Rev. B}\ }\textbf {\bibinfo {volume} {89}},\ \bibinfo {pages} {224408}
  (\bibinfo {year} {2016})}\BibitemShut {NoStop}%
\bibitem [{\citenamefont {Kim}\ \emph {et~al.}(2016)\citenamefont {Kim},
  \citenamefont {Stamps},\ and\ \citenamefont {Camley}}]{Kim16}%
  \BibitemOpen
  \bibfield  {author} {\bibinfo {author} {\bibfnamefont {J.-V.}\ \bibnamefont
  {Kim}}, \bibinfo {author} {\bibfnamefont {R.~L.}\ \bibnamefont {Stamps}}, \
  and\ \bibinfo {author} {\bibfnamefont {R.~E.}\ \bibnamefont {Camley}},\
  }\href@noop {} {\bibfield  {journal} {\bibinfo  {journal} {Phys. Rev. Lett}\
  }\textbf {\bibinfo {volume} {117}},\ \bibinfo {pages} {197204} (\bibinfo
  {year} {2016})}\BibitemShut {NoStop}%
\bibitem [{\citenamefont {Borys}\ \emph {et~al.}(2016)\citenamefont {Borys},
  \citenamefont {Garcia-Sanchez}, \citenamefont {Kim},\ and\ \citenamefont
  {Stamps}}]{Borys16}%
  \BibitemOpen
  \bibfield  {author} {\bibinfo {author} {\bibfnamefont {P.}~\bibnamefont
  {Borys}}, \bibinfo {author} {\bibfnamefont {F.}~\bibnamefont
  {Garcia-Sanchez}}, \bibinfo {author} {\bibfnamefont {J.-V.}\ \bibnamefont
  {Kim}}, \ and\ \bibinfo {author} {\bibfnamefont {R.~L.}\ \bibnamefont
  {Stamps}},\ }\href {\doibase https://doi.org/10.1002/aelm.201500202}
  {\bibfield  {journal} {\bibinfo  {journal} {Adv. Electron.Mater.}\ }\textbf
  {\bibinfo {volume} {2}},\ \bibinfo {pages} {1500202} (\bibinfo {year}
  {2016})}\BibitemShut {NoStop}%
\bibitem [{\citenamefont {Whitehead}\ \emph {et~al.}(2017)\citenamefont
  {Whitehead}, \citenamefont {Horsley}, \citenamefont {Philbin}, \citenamefont
  {Kuchko},\ and\ \citenamefont {Kruglyak}}]{Whitehead17}%
  \BibitemOpen
  \bibfield  {author} {\bibinfo {author} {\bibfnamefont {N.~J.}\ \bibnamefont
  {Whitehead}}, \bibinfo {author} {\bibfnamefont {S.~A.~R.}\ \bibnamefont
  {Horsley}}, \bibinfo {author} {\bibfnamefont {T.~G.}\ \bibnamefont
  {Philbin}}, \bibinfo {author} {\bibfnamefont {A.~N.}\ \bibnamefont {Kuchko}},
  \ and\ \bibinfo {author} {\bibfnamefont {V.~V.}\ \bibnamefont {Kruglyak}},\
  }\href@noop {} {\bibfield  {journal} {\bibinfo  {journal} {Phys. Rev. B}\
  }\textbf {\bibinfo {volume} {96}},\ \bibinfo {pages} {064415} (\bibinfo
  {year} {2017})}\BibitemShut {NoStop}%
\bibitem [{\citenamefont {Zingsem}\ \emph {et~al.}(2019)\citenamefont
  {Zingsem}, \citenamefont {Farle}, \citenamefont {Stamps},\ and\ \citenamefont
  {Camley}}]{Zingsem19}%
  \BibitemOpen
  \bibfield  {author} {\bibinfo {author} {\bibfnamefont {B.~W.}\ \bibnamefont
  {Zingsem}}, \bibinfo {author} {\bibfnamefont {M.}~\bibnamefont {Farle}},
  \bibinfo {author} {\bibfnamefont {R.~L.}\ \bibnamefont {Stamps}}, \ and\
  \bibinfo {author} {\bibfnamefont {R.~E.}\ \bibnamefont {Camley}},\
  }\href@noop {} {\bibfield  {journal} {\bibinfo  {journal} {Phys. Rev. B}\
  }\textbf {\bibinfo {volume} {99}},\ \bibinfo {pages} {214429} (\bibinfo
  {year} {2019})}\BibitemShut {NoStop}%
\bibitem [{\citenamefont {Togawa}\ \emph {et~al.}(2012)\citenamefont {Togawa},
  \citenamefont {Koyama}, \citenamefont {Takayanagi}, \citenamefont {Mori},
  \citenamefont {Kousaka}, \citenamefont {Akimitsu}, \citenamefont {Nishihara},
  \citenamefont {Inoue}, \citenamefont {Ovchinnikov},\ and\ \citenamefont
  {Kishine}}]{Togawa12}%
  \BibitemOpen
  \bibfield  {author} {\bibinfo {author} {\bibfnamefont {Y.}~\bibnamefont
  {Togawa}}, \bibinfo {author} {\bibfnamefont {T.}~\bibnamefont {Koyama}},
  \bibinfo {author} {\bibfnamefont {T.}~\bibnamefont {Takayanagi}}, \bibinfo
  {author} {\bibfnamefont {S.}~\bibnamefont {Mori}}, \bibinfo {author}
  {\bibfnamefont {Y.}~\bibnamefont {Kousaka}}, \bibinfo {author} {\bibfnamefont
  {J.}~\bibnamefont {Akimitsu}}, \bibinfo {author} {\bibfnamefont
  {S.}~\bibnamefont {Nishihara}}, \bibinfo {author} {\bibfnamefont
  {K.}~\bibnamefont {Inoue}}, \bibinfo {author} {\bibfnamefont
  {A.}~\bibnamefont {Ovchinnikov}}, \ and\ \bibinfo {author} {\bibfnamefont
  {J.}~\bibnamefont {Kishine}},\ }\href@noop {} {\bibfield  {journal} {\bibinfo
   {journal} {Phys. Rev. Lett.}\ }\textbf {\bibinfo {volume} {108}},\ \bibinfo
  {pages} {107202} (\bibinfo {year} {2012})}\BibitemShut {NoStop}%
\bibitem [{\citenamefont {Laliena}\ \emph
  {et~al.}(2016{\natexlab{a}})\citenamefont {Laliena}, \citenamefont {Campo},
  \citenamefont {Kishine}, \citenamefont {Ovchinnikov}, \citenamefont {Togawa},
  \citenamefont {Kousaka},\ and\ \citenamefont {Inoue}}]{Laliena16a}%
  \BibitemOpen
  \bibfield  {author} {\bibinfo {author} {\bibfnamefont {V.}~\bibnamefont
  {Laliena}}, \bibinfo {author} {\bibfnamefont {J.}~\bibnamefont {Campo}},
  \bibinfo {author} {\bibfnamefont {J.}~\bibnamefont {Kishine}}, \bibinfo
  {author} {\bibfnamefont {A.}~\bibnamefont {Ovchinnikov}}, \bibinfo {author}
  {\bibfnamefont {Y.}~\bibnamefont {Togawa}}, \bibinfo {author} {\bibfnamefont
  {Y.}~\bibnamefont {Kousaka}}, \ and\ \bibinfo {author} {\bibfnamefont
  {K.}~\bibnamefont {Inoue}},\ }\href@noop {} {\bibfield  {journal} {\bibinfo
  {journal} {Phys. Rev. B}\ }\textbf {\bibinfo {volume} {93}},\ \bibinfo
  {pages} {134424} (\bibinfo {year} {2016}{\natexlab{a}})}\BibitemShut
  {NoStop}%
\bibitem [{\citenamefont {Laliena}\ \emph
  {et~al.}(2016{\natexlab{b}})\citenamefont {Laliena}, \citenamefont {Campo},\
  and\ \citenamefont {Kousaka}}]{Laliena16b}%
  \BibitemOpen
  \bibfield  {author} {\bibinfo {author} {\bibfnamefont {V.}~\bibnamefont
  {Laliena}}, \bibinfo {author} {\bibfnamefont {J.}~\bibnamefont {Campo}}, \
  and\ \bibinfo {author} {\bibfnamefont {Y.}~\bibnamefont {Kousaka}},\
  }\href@noop {} {\bibfield  {journal} {\bibinfo  {journal} {Phys. Rev. B}\
  }\textbf {\bibinfo {volume} {94}},\ \bibinfo {pages} {094439} (\bibinfo
  {year} {2016}{\natexlab{b}})}\BibitemShut {NoStop}%
\bibitem [{\citenamefont {Shinozaki}\ \emph {et~al.}(2016)\citenamefont
  {Shinozaki}, \citenamefont {Hoshino}, \citenamefont {Masaki}, \citenamefont
  {Kishine},\ and\ \citenamefont {Kato}}]{Shinozaki16}%
  \BibitemOpen
  \bibfield  {author} {\bibinfo {author} {\bibfnamefont {M.}~\bibnamefont
  {Shinozaki}}, \bibinfo {author} {\bibfnamefont {S.}~\bibnamefont {Hoshino}},
  \bibinfo {author} {\bibfnamefont {Y.}~\bibnamefont {Masaki}}, \bibinfo
  {author} {\bibfnamefont {J.}~\bibnamefont {Kishine}}, \ and\ \bibinfo
  {author} {\bibfnamefont {Y.}~\bibnamefont {Kato}},\ }\href@noop {} {\bibfield
   {journal} {\bibinfo  {journal} {J. Phys. Soc. Jpn. 85}\ ,\ \bibinfo {pages}
  {074710}} (\bibinfo {year} {2016})}\BibitemShut {NoStop}%
\bibitem [{\citenamefont {Tsuruta}\ \emph {et~al.}(2016)\citenamefont
  {Tsuruta}, \citenamefont {Mito}, \citenamefont {Deguchi}, \citenamefont
  {Kishine}, \citenamefont {Kousaka}, \citenamefont {Akimitsu},\ and\
  \citenamefont {Inoue}}]{Tsuruta16}%
  \BibitemOpen
  \bibfield  {author} {\bibinfo {author} {\bibfnamefont {K.}~\bibnamefont
  {Tsuruta}}, \bibinfo {author} {\bibfnamefont {M.}~\bibnamefont {Mito}},
  \bibinfo {author} {\bibfnamefont {H.}~\bibnamefont {Deguchi}}, \bibinfo
  {author} {\bibfnamefont {J.}~\bibnamefont {Kishine}}, \bibinfo {author}
  {\bibfnamefont {Y.}~\bibnamefont {Kousaka}}, \bibinfo {author} {\bibfnamefont
  {J.}~\bibnamefont {Akimitsu}}, \ and\ \bibinfo {author} {\bibfnamefont
  {K.}~\bibnamefont {Inoue}},\ }\href@noop {} {\bibfield  {journal} {\bibinfo
  {journal} {Phys. Rev. B}\ }\textbf {\bibinfo {volume} {93}},\ \bibinfo
  {pages} {104402} (\bibinfo {year} {2016})}\BibitemShut {NoStop}%
\bibitem [{\citenamefont {Kishine}\ \emph {et~al.}(2016)\citenamefont
  {Kishine}, \citenamefont {Proskurin}, \citenamefont {Bostrem}, \citenamefont
  {Ovchinnikov},\ and\ \citenamefont {Sinitsyn}}]{Kishine16}%
  \BibitemOpen
  \bibfield  {author} {\bibinfo {author} {\bibfnamefont {J.}~\bibnamefont
  {Kishine}}, \bibinfo {author} {\bibfnamefont {I.}~\bibnamefont {Proskurin}},
  \bibinfo {author} {\bibfnamefont {I.~G.}\ \bibnamefont {Bostrem}}, \bibinfo
  {author} {\bibfnamefont {A.~S.}\ \bibnamefont {Ovchinnikov}}, \ and\ \bibinfo
  {author} {\bibfnamefont {V.~E.}\ \bibnamefont {Sinitsyn}},\ }\href@noop {}
  {\bibfield  {journal} {\bibinfo  {journal} {Phys. Rev. B}\ }\textbf {\bibinfo
  {volume} {93}},\ \bibinfo {pages} {054403} (\bibinfo {year}
  {2016})}\BibitemShut {NoStop}%
\bibitem [{\citenamefont {Laliena}\ \emph {et~al.}(2017)\citenamefont
  {Laliena}, \citenamefont {Campo},\ and\ \citenamefont
  {Kousaka}}]{Laliena17a}%
  \BibitemOpen
  \bibfield  {author} {\bibinfo {author} {\bibfnamefont {V.}~\bibnamefont
  {Laliena}}, \bibinfo {author} {\bibfnamefont {J.}~\bibnamefont {Campo}}, \
  and\ \bibinfo {author} {\bibfnamefont {Y.}~\bibnamefont {Kousaka}},\
  }\href@noop {} {\bibfield  {journal} {\bibinfo  {journal} {Phys. Rev. B}\
  }\textbf {\bibinfo {volume} {95}},\ \bibinfo {pages} {224410} (\bibinfo
  {year} {2017})}\BibitemShut {NoStop}%
\bibitem [{\citenamefont {Goncalves}\ \emph {et~al.}(2017)\citenamefont
  {Goncalves}, \citenamefont {Sogo}, \citenamefont {Shimamoto}, \citenamefont
  {Kousaka}, \citenamefont {Akimitsu}, \citenamefont {Nishihara}, \citenamefont
  {Inoue}, \citenamefont {Yoshizawa}, \citenamefont {Hagiwara}, \citenamefont
  {Mito}, \citenamefont {Stamps}, \citenamefont {Bostrem}, \citenamefont
  {Sinitsyn}, \citenamefont {Ovchinnikov}, \citenamefont {Kishine},\ and\
  \citenamefont {Togawa}}]{Goncalves17}%
  \BibitemOpen
  \bibfield  {author} {\bibinfo {author} {\bibfnamefont {F.~J.~T.}\
  \bibnamefont {Goncalves}}, \bibinfo {author} {\bibfnamefont {T.}~\bibnamefont
  {Sogo}}, \bibinfo {author} {\bibfnamefont {Y.}~\bibnamefont {Shimamoto}},
  \bibinfo {author} {\bibfnamefont {Y.}~\bibnamefont {Kousaka}}, \bibinfo
  {author} {\bibfnamefont {J.}~\bibnamefont {Akimitsu}}, \bibinfo {author}
  {\bibfnamefont {S.}~\bibnamefont {Nishihara}}, \bibinfo {author}
  {\bibfnamefont {K.}~\bibnamefont {Inoue}}, \bibinfo {author} {\bibfnamefont
  {D.}~\bibnamefont {Yoshizawa}}, \bibinfo {author} {\bibfnamefont
  {M.}~\bibnamefont {Hagiwara}}, \bibinfo {author} {\bibfnamefont
  {M.}~\bibnamefont {Mito}}, \bibinfo {author} {\bibfnamefont {R.~L.}\
  \bibnamefont {Stamps}}, \bibinfo {author} {\bibfnamefont {I.~G.}\
  \bibnamefont {Bostrem}}, \bibinfo {author} {\bibfnamefont {V.~E.}\
  \bibnamefont {Sinitsyn}}, \bibinfo {author} {\bibfnamefont {A.~S.}\
  \bibnamefont {Ovchinnikov}}, \bibinfo {author} {\bibfnamefont
  {J.}~\bibnamefont {Kishine}}, \ and\ \bibinfo {author} {\bibfnamefont
  {Y.}~\bibnamefont {Togawa}},\ }\href@noop {} {\bibfield  {journal} {\bibinfo
  {journal} {Phys. Rev. B}\ }\textbf {\bibinfo {volume} {95}},\ \bibinfo
  {pages} {104415} (\bibinfo {year} {2017})}\BibitemShut {NoStop}%
\bibitem [{\citenamefont {Laliena}\ \emph {et~al.}(2018)\citenamefont
  {Laliena}, \citenamefont {Albalate},\ and\ \citenamefont
  {Campo}}]{Laliena18c}%
  \BibitemOpen
  \bibfield  {author} {\bibinfo {author} {\bibfnamefont {V.}~\bibnamefont
  {Laliena}}, \bibinfo {author} {\bibfnamefont {G.}~\bibnamefont {Albalate}}, \
  and\ \bibinfo {author} {\bibfnamefont {J.}~\bibnamefont {Campo}},\
  }\href@noop {} {\bibfield  {journal} {\bibinfo  {journal} {Phys. Rev. B}\
  }\textbf {\bibinfo {volume} {98}},\ \bibinfo {pages} {224407} (\bibinfo
  {year} {2018})}\BibitemShut {NoStop}%
\bibitem [{\citenamefont {Masaki}\ \emph {et~al.}(2018)\citenamefont {Masaki},
  \citenamefont {Aoki}, \citenamefont {Togawa},\ and\ \citenamefont
  {Kato}}]{Masaki18}%
  \BibitemOpen
  \bibfield  {author} {\bibinfo {author} {\bibfnamefont {Y.}~\bibnamefont
  {Masaki}}, \bibinfo {author} {\bibfnamefont {R.}~\bibnamefont {Aoki}},
  \bibinfo {author} {\bibfnamefont {Y.}~\bibnamefont {Togawa}}, \ and\ \bibinfo
  {author} {\bibfnamefont {Y.}~\bibnamefont {Kato}},\ }\href@noop {} {\bibfield
   {journal} {\bibinfo  {journal} {Phys. Rev. B}\ }\textbf {\bibinfo {volume}
  {98}},\ \bibinfo {pages} {100402(R)} (\bibinfo {year} {2018})}\BibitemShut
  {NoStop}%
\bibitem [{\citenamefont {Kishine}\ and\ \citenamefont
  {Ovchinnikov}(2020)}]{Kishine20}%
  \BibitemOpen
  \bibfield  {author} {\bibinfo {author} {\bibfnamefont {J.}~\bibnamefont
  {Kishine}}\ and\ \bibinfo {author} {\bibfnamefont {A.~S.}\ \bibnamefont
  {Ovchinnikov}},\ }\href {\doibase 10.1103/PhysRevB.101.184425} {\bibfield
  {journal} {\bibinfo  {journal} {Phys. Rev. B}\ }\textbf {\bibinfo {volume}
  {101}},\ \bibinfo {pages} {184425} (\bibinfo {year} {2020})}\BibitemShut
  {NoStop}%
\bibitem [{\citenamefont {Laliena}\ \emph {et~al.}(2020)\citenamefont
  {Laliena}, \citenamefont {Bustingorry},\ and\ \citenamefont
  {Campo}}]{Laliena20}%
  \BibitemOpen
  \bibfield  {author} {\bibinfo {author} {\bibfnamefont {V.}~\bibnamefont
  {Laliena}}, \bibinfo {author} {\bibfnamefont {S.}~\bibnamefont
  {Bustingorry}}, \ and\ \bibinfo {author} {\bibfnamefont {J.}~\bibnamefont
  {Campo}},\ }\href {\doibase https://doi.org/10.1038/s41598-020-76903-8}
  {\bibfield  {journal} {\bibinfo  {journal} {Sci Rep}\ } (\bibinfo {year}
  {2020}),\ https://doi.org/10.1038/s41598-020-76903-8}\BibitemShut {NoStop}%
\bibitem [{Note1()}]{Note1}%
  \BibitemOpen
  \bibinfo {note} {To avoid any misinterpretation, let us clarify that the term
  ``spinor'' is used here to distinguish the two-component object $\protect
  \mathaccentV {tilde}07E{\xi }$ from one-component fields and
  three-dimensional vectors. Obviously, it is an abuse of language, since
  spinors are related to spatial rotations in a very precise way, very
  different from our $\protect \mathaccentV {tilde}07E{\xi }$.}\BibitemShut
  {Stop}%
\bibitem [{Note2()}]{Note2}%
  \BibitemOpen
  \bibinfo {note} {If $K$ has zero modes, $K^{-1/2}$ is not defined, and this
  argument is problematic, but it could be modified to circumvent the
  problem}\BibitemShut {NoStop}%
\bibitem [{\citenamefont {Laliena}\ and\ \citenamefont
  {Campo}(2017)}]{Laliena17b}%
  \BibitemOpen
  \bibfield  {author} {\bibinfo {author} {\bibfnamefont {V.}~\bibnamefont
  {Laliena}}\ and\ \bibinfo {author} {\bibfnamefont {J.}~\bibnamefont
  {Campo}},\ }\href@noop {} {\bibfield  {journal} {\bibinfo  {journal} {Phys.
  Rev. B}\ }\textbf {\bibinfo {volume} {96}},\ \bibinfo {pages} {134420}
  (\bibinfo {year} {2017})}\BibitemShut {NoStop}%
\bibitem [{Note3()}]{Note3}%
  \BibitemOpen
  \bibinfo {note} {These results can be easily obtained by noticing that
  $\Omega _2\Omega _1$ is a hermitian positive (semi)definite operator with
  respect to the scalar product \unhbox \voidb@x \hbox {$((f,g)) = (f,\omega
  _0\Omega _2^{-1}g)$}. Therefore, the eigenvalues of $\Omega _2\Omega _1$ are
  real and non-negative and its eigenfunctions are ortohogonal with respect to
  the $((,))$ product, what amounts to Eq.~(\ref {eq:norm1}).}\BibitemShut
  {Stop}%
\bibitem [{\citenamefont {Dzyaloshinskii}(1964)}]{Dzyal64}%
  \BibitemOpen
  \bibfield  {author} {\bibinfo {author} {\bibfnamefont {I.}~\bibnamefont
  {Dzyaloshinskii}},\ }\href@noop {} {\bibfield  {journal} {\bibinfo  {journal}
  {Sov. Phys. JETP}\ }\textbf {\bibinfo {volume} {19}},\ \bibinfo {pages} {960}
  (\bibinfo {year} {1964})}\BibitemShut {NoStop}%
\bibitem [{\citenamefont {Laliena}\ and\ \citenamefont {Campo}(2020)}]{suppl}%
  \BibitemOpen
  \bibfield  {author} {\bibinfo {author} {\bibfnamefont {V.}~\bibnamefont
  {Laliena}}\ and\ \bibinfo {author} {\bibfnamefont {J.}~\bibnamefont
  {Campo}},\ }\href@noop {} {\bibfield  {journal} {\bibinfo  {journal}
  {Supplemental material to this article}\ } (\bibinfo {year}
  {2020})}\BibitemShut {NoStop}%
\bibitem [{\citenamefont {Galindo}\ and\ \citenamefont
  {Pascual}(1990)}]{Galindo90I}%
  \BibitemOpen
  \bibfield  {author} {\bibinfo {author} {\bibfnamefont {A.}~\bibnamefont
  {Galindo}}\ and\ \bibinfo {author} {\bibfnamefont {P.}~\bibnamefont
  {Pascual}},\ }\href@noop {} {\emph {\bibinfo {title} {Quantum Mechanics}}},\
  Vol.~\bibinfo {volume} {I}\ (\bibinfo  {publisher} {Springer-Verlag},\
  \bibinfo {year} {1990})\BibitemShut {NoStop}%
\bibitem [{\citenamefont {Levinson}(1949)}]{Levinson49}%
  \BibitemOpen
  \bibfield  {author} {\bibinfo {author} {\bibfnamefont {N.}~\bibnamefont
  {Levinson}},\ }\href@noop {} {\bibfield  {journal} {\bibinfo  {journal} {Kgl.
  Danske. Videnskab. Selskab., Mat.-Fys. Medd.}\ }\textbf {\bibinfo {volume}
  {25}} (\bibinfo {year} {1949})}\BibitemShut {NoStop}%
\bibitem [{\citenamefont {Wigner}(1955)}]{Wigner55}%
  \BibitemOpen
  \bibfield  {author} {\bibinfo {author} {\bibfnamefont {E.}~\bibnamefont
  {Wigner}},\ }\href {\doibase 10.1103/PhysRev.98.145} {\bibfield  {journal}
  {\bibinfo  {journal} {Phys. Rev.}\ }\textbf {\bibinfo {volume} {98}},\
  \bibinfo {pages} {145} (\bibinfo {year} {1955})}\BibitemShut {NoStop}%
\bibitem [{\citenamefont {Artmann}(1948)}]{Artmann48}%
  \BibitemOpen
  \bibfield  {author} {\bibinfo {author} {\bibfnamefont {K.}~\bibnamefont
  {Artmann}},\ }\href {\doibase 10.1002/andp.19484370108} {\bibfield  {journal}
  {\bibinfo  {journal} {Annalen der Physik}\ }\textbf {\bibinfo {volume}
  {437}},\ \bibinfo {pages} {87} (\bibinfo {year} {1948})}\BibitemShut
  {NoStop}%
\bibitem [{\citenamefont {Goos}\ and\ \citenamefont
  {H\"anchen}(1947)}]{Goos47}%
  \BibitemOpen
  \bibfield  {author} {\bibinfo {author} {\bibfnamefont {F.}~\bibnamefont
  {Goos}}\ and\ \bibinfo {author} {\bibfnamefont {H.}~\bibnamefont
  {H\"anchen}},\ }\href {\doibase 10.1002/andp.19474360704} {\bibfield
  {journal} {\bibinfo  {journal} {Ann. Phys.}\ }\textbf {\bibinfo {volume}
  {436}},\ \bibinfo {pages} {333} (\bibinfo {year} {1947})}\BibitemShut
  {NoStop}%
\bibitem [{\citenamefont {Dadoenkova}\ \emph {et~al.}(2012)\citenamefont
  {Dadoenkova}, \citenamefont {Dadoenkova}, \citenamefont {Lyubchanskii},
  \citenamefont {Sokolovskyy}, \citenamefont {Kłos}, \citenamefont
  {Romero-Vivas},\ and\ \citenamefont {Krawczyk}}]{Dadoenkova12}%
  \BibitemOpen
  \bibfield  {author} {\bibinfo {author} {\bibfnamefont {Y.~S.}\ \bibnamefont
  {Dadoenkova}}, \bibinfo {author} {\bibfnamefont {N.~N.}\ \bibnamefont
  {Dadoenkova}}, \bibinfo {author} {\bibfnamefont {I.~L.}\ \bibnamefont
  {Lyubchanskii}}, \bibinfo {author} {\bibfnamefont {M.~L.}\ \bibnamefont
  {Sokolovskyy}}, \bibinfo {author} {\bibfnamefont {J.~W.}\ \bibnamefont
  {Kłos}}, \bibinfo {author} {\bibfnamefont {J.}~\bibnamefont {Romero-Vivas}},
  \ and\ \bibinfo {author} {\bibfnamefont {M.}~\bibnamefont {Krawczyk}},\
  }\href {\doibase 10.1063/1.4738987} {\bibfield  {journal} {\bibinfo
  {journal} {Appl. Phys. Lett.}\ }\textbf {\bibinfo {volume} {101}},\ \bibinfo
  {pages} {042404} (\bibinfo {year} {2012})}\BibitemShut {NoStop}%
\bibitem [{\citenamefont {Gruszecki}\ \emph {et~al.}(2014)\citenamefont
  {Gruszecki}, \citenamefont {Romero-Vivas}, \citenamefont {Dadoenkova},
  \citenamefont {Dadoenkova}, \citenamefont {Lyubchanskii},\ and\ \citenamefont
  {Krawczyk}}]{Gruszecki14}%
  \BibitemOpen
  \bibfield  {author} {\bibinfo {author} {\bibfnamefont {P.}~\bibnamefont
  {Gruszecki}}, \bibinfo {author} {\bibfnamefont {J.}~\bibnamefont
  {Romero-Vivas}}, \bibinfo {author} {\bibfnamefont {Y.~S.}\ \bibnamefont
  {Dadoenkova}}, \bibinfo {author} {\bibfnamefont {N.~N.}\ \bibnamefont
  {Dadoenkova}}, \bibinfo {author} {\bibfnamefont {I.~L.}\ \bibnamefont
  {Lyubchanskii}}, \ and\ \bibinfo {author} {\bibfnamefont {M.}~\bibnamefont
  {Krawczyk}},\ }\href {\doibase 10.1063/1.4904342} {\bibfield  {journal}
  {\bibinfo  {journal} {Appl. Phys. Lett.}\ }\textbf {\bibinfo {volume}
  {105}},\ \bibinfo {pages} {242406} (\bibinfo {year} {2014})}\BibitemShut
  {NoStop}%
\bibitem [{\citenamefont {Gruszecki}\ \emph {et~al.}(2015)\citenamefont
  {Gruszecki}, \citenamefont {Dadoenkova}, \citenamefont {Dadoenkova},
  \citenamefont {Lyubchanskii}, \citenamefont {Romero-Vivas}, \citenamefont
  {Guslienko},\ and\ \citenamefont {Krawczyk}}]{Gruszecki15}%
  \BibitemOpen
  \bibfield  {author} {\bibinfo {author} {\bibfnamefont {P.}~\bibnamefont
  {Gruszecki}}, \bibinfo {author} {\bibfnamefont {Y.~S.}\ \bibnamefont
  {Dadoenkova}}, \bibinfo {author} {\bibfnamefont {N.~N.}\ \bibnamefont
  {Dadoenkova}}, \bibinfo {author} {\bibfnamefont {I.~L.}\ \bibnamefont
  {Lyubchanskii}}, \bibinfo {author} {\bibfnamefont {J.}~\bibnamefont
  {Romero-Vivas}}, \bibinfo {author} {\bibfnamefont {K.~Y.}\ \bibnamefont
  {Guslienko}}, \ and\ \bibinfo {author} {\bibfnamefont {M.}~\bibnamefont
  {Krawczyk}},\ }\href {\doibase 10.1103/PhysRevB.92.054427} {\bibfield
  {journal} {\bibinfo  {journal} {Phys. Rev. B}\ }\textbf {\bibinfo {volume}
  {92}},\ \bibinfo {pages} {054427} (\bibinfo {year} {2015})}\BibitemShut
  {NoStop}%
\bibitem [{\citenamefont {Gruszecki}\ \emph {et~al.}(2017)\citenamefont
  {Gruszecki}, \citenamefont {Mailyan}, \citenamefont {Gorobets},\ and\
  \citenamefont {Krawczyk}}]{Gruszecki17}%
  \BibitemOpen
  \bibfield  {author} {\bibinfo {author} {\bibfnamefont {P.}~\bibnamefont
  {Gruszecki}}, \bibinfo {author} {\bibfnamefont {M.}~\bibnamefont {Mailyan}},
  \bibinfo {author} {\bibfnamefont {O.}~\bibnamefont {Gorobets}}, \ and\
  \bibinfo {author} {\bibfnamefont {M.}~\bibnamefont {Krawczyk}},\ }\href
  {\doibase 10.1103/PhysRevB.95.014421} {\bibfield  {journal} {\bibinfo
  {journal} {Phys. Rev. B}\ }\textbf {\bibinfo {volume} {95}},\ \bibinfo
  {pages} {014421} (\bibinfo {year} {2017})}\BibitemShut {NoStop}%
\bibitem [{\citenamefont {{Mailyan}}\ \emph {et~al.}(2017)\citenamefont
  {{Mailyan}}, \citenamefont {{Gruszecki}}, \citenamefont {{Gorobets}},\ and\
  \citenamefont {{Krawczyk}}}]{Mailyan17}%
  \BibitemOpen
  \bibfield  {author} {\bibinfo {author} {\bibfnamefont {M.}~\bibnamefont
  {{Mailyan}}}, \bibinfo {author} {\bibfnamefont {P.}~\bibnamefont
  {{Gruszecki}}}, \bibinfo {author} {\bibfnamefont {O.}~\bibnamefont
  {{Gorobets}}}, \ and\ \bibinfo {author} {\bibfnamefont {M.}~\bibnamefont
  {{Krawczyk}}},\ }\href {\doibase 10.1109/TMAG.2017.2696581} {\bibfield
  {journal} {\bibinfo  {journal} {IEEE Transactions on Magnetics}\ }\textbf
  {\bibinfo {volume} {53}},\ \bibinfo {pages} {1} (\bibinfo {year}
  {2017})}\BibitemShut {NoStop}%
\bibitem [{\citenamefont {Wang}\ \emph {et~al.}(2019)\citenamefont {Wang},
  \citenamefont {Cao},\ and\ \citenamefont {Yan}}]{Wang19}%
  \BibitemOpen
  \bibfield  {author} {\bibinfo {author} {\bibfnamefont {Z.}~\bibnamefont
  {Wang}}, \bibinfo {author} {\bibfnamefont {Y.}~\bibnamefont {Cao}}, \ and\
  \bibinfo {author} {\bibfnamefont {P.}~\bibnamefont {Yan}},\ }\href {\doibase
  10.1103/PhysRevB.100.064421} {\bibfield  {journal} {\bibinfo  {journal}
  {Phys. Rev. B}\ }\textbf {\bibinfo {volume} {100}},\ \bibinfo {pages}
  {064421} (\bibinfo {year} {2019})}\BibitemShut {NoStop}%
\bibitem [{\citenamefont {Zhen}\ and\ \citenamefont {Deng}(2020)}]{Zhen20}%
  \BibitemOpen
  \bibfield  {author} {\bibinfo {author} {\bibfnamefont {W.}~\bibnamefont
  {Zhen}}\ and\ \bibinfo {author} {\bibfnamefont {D.}~\bibnamefont {Deng}},\
  }\href {\doibase https://doi.org/10.1016/j.optcom.2020.126067} {\bibfield
  {journal} {\bibinfo  {journal} {Optics Communications}\ }\textbf {\bibinfo
  {volume} {474}},\ \bibinfo {pages} {126067} (\bibinfo {year}
  {2020})}\BibitemShut {NoStop}%
\bibitem [{\citenamefont {Stigloher}\ \emph {et~al.}(2018)\citenamefont
  {Stigloher}, \citenamefont {Taniguchi}, \citenamefont {K\"orner},
  \citenamefont {Decker}, \citenamefont {Moriyama}, \citenamefont {Ono},\ and\
  \citenamefont {Back}}]{Stigloher18}%
  \BibitemOpen
  \bibfield  {author} {\bibinfo {author} {\bibfnamefont {J.}~\bibnamefont
  {Stigloher}}, \bibinfo {author} {\bibfnamefont {T.}~\bibnamefont
  {Taniguchi}}, \bibinfo {author} {\bibfnamefont {H.~S.}\ \bibnamefont
  {K\"orner}}, \bibinfo {author} {\bibfnamefont {M.}~\bibnamefont {Decker}},
  \bibinfo {author} {\bibfnamefont {T.}~\bibnamefont {Moriyama}}, \bibinfo
  {author} {\bibfnamefont {T.}~\bibnamefont {Ono}}, \ and\ \bibinfo {author}
  {\bibfnamefont {C.~H.}\ \bibnamefont {Back}},\ }\href {\doibase
  10.1103/PhysRevLett.121.137201} {\bibfield  {journal} {\bibinfo  {journal}
  {Phys. Rev. Lett.}\ }\textbf {\bibinfo {volume} {121}},\ \bibinfo {pages}
  {137201} (\bibinfo {year} {2018})}\BibitemShut {NoStop}%
\end{thebibliography}%


\begin{thebibliography}{3}%
\makeatletter
\providecommand \@ifxundefined [1]{%
 \@ifx{#1\undefined}
}%
\providecommand \@ifnum [1]{%
 \ifnum #1\expandafter \@firstoftwo
 \else \expandafter \@secondoftwo
 \fi
}%
\providecommand \@ifx [1]{%
 \ifx #1\expandafter \@firstoftwo
 \else \expandafter \@secondoftwo
 \fi
}%
\providecommand \natexlab [1]{#1}%
\providecommand \enquote  [1]{``#1''}%
\providecommand \bibnamefont  [1]{#1}%
\providecommand \bibfnamefont [1]{#1}%
\providecommand \citenamefont [1]{#1}%
\providecommand \href@noop [0]{\@secondoftwo}%
\providecommand \href [0]{\begingroup \@sanitize@url \@href}%
\providecommand \@href[1]{\@@startlink{#1}\@@href}%
\providecommand \@@href[1]{\endgroup#1\@@endlink}%
\providecommand \@sanitize@url [0]{\catcode `\\12\catcode `\$12\catcode
  `\&12\catcode `\#12\catcode `\^12\catcode `\_12\catcode `\%12\relax}%
\providecommand \@@startlink[1]{}%
\providecommand \@@endlink[0]{}%
\providecommand \url  [0]{\begingroup\@sanitize@url \@url }%
\providecommand \@url [1]{\endgroup\@href {#1}{\urlprefix }}%
\providecommand \urlprefix  [0]{URL }%
\providecommand \Eprint [0]{\href }%
\providecommand \doibase [0]{http://dx.doi.org/}%
\providecommand \selectlanguage [0]{\@gobble}%
\providecommand \bibinfo  [0]{\@secondoftwo}%
\providecommand \bibfield  [0]{\@secondoftwo}%
\providecommand \translation [1]{[#1]}%
\providecommand \BibitemOpen [0]{}%
\providecommand \bibitemStop [0]{}%
\providecommand \bibitemNoStop [0]{.\EOS\space}%
\providecommand \EOS [0]{\spacefactor3000\relax}%
\providecommand \BibitemShut  [1]{\csname bibitem#1\endcsname}%
\let\auto@bib@innerbib\@empty
\bibitem [{\citenamefont {Lehoucq}\ \emph {et~al.}(1998)\citenamefont
  {Lehoucq}, \citenamefont {Sorensen},\ and\ \citenamefont {Yang}}]{ARPACK}%
  \BibitemOpen
  \bibfield  {author} {\bibinfo {author} {\bibfnamefont {R.}~\bibnamefont
  {Lehoucq}}, \bibinfo {author} {\bibfnamefont {D.}~\bibnamefont {Sorensen}}, \
  and\ \bibinfo {author} {\bibfnamefont {C.}~\bibnamefont {Yang}},\ }\href@noop
  {} {\emph {\bibinfo {title} {ARPACK Users Guide: Solution of Large-Scale
  Eigenvalue Problems with Implicitly Restarted Arnoldi Methods.}}}\ (\bibinfo
  {publisher} {SIAM},\ \bibinfo {address} {Philadelphia},\ \bibinfo {year}
  {1998})\BibitemShut {NoStop}%
\bibitem [{\citenamefont {Galindo}\ and\ \citenamefont
  {Pascual}(1990)}]{Galindo90I}%
  \BibitemOpen
  \bibfield  {author} {\bibinfo {author} {\bibfnamefont {A.}~\bibnamefont
  {Galindo}}\ and\ \bibinfo {author} {\bibfnamefont {P.}~\bibnamefont
  {Pascual}},\ }\href@noop {} {\emph {\bibinfo {title} {Quantum Mechanics}}},\
  Vol.~\bibinfo {volume} {I}\ (\bibinfo  {publisher} {Springer-Verlag},\
  \bibinfo {year} {1990})\BibitemShut {NoStop}%
\bibitem [{\citenamefont {Levinson}(1949)}]{Levinson49}%
  \BibitemOpen
  \bibfield  {author} {\bibinfo {author} {\bibfnamefont {N.}~\bibnamefont
  {Levinson}},\ }\href@noop {} {\bibfield  {journal} {\bibinfo  {journal} {Kgl.
  Danske. Videnskab. Selskab., Mat.-Fys. Medd.}\ }\textbf {\bibinfo {volume}
  {25}} (\bibinfo {year} {1949})}\BibitemShut {NoStop}%
\end{thebibliography}%

%

\end{document}


\title{Supplemental material to
  ``Magnonic Goos-H\"anchen effect induced by one dimensional solitons''}

\author{Victor Laliena}
\email[]{laliena@unizar.es}
\affiliation{Aragon Nanoscience and Materials Institute 
  (CSIC -- University of Zaragoza) and
  Condensed Matter Physics Department, University of Zaragoza \\
  C/Pedro Cerbuna 12, Zaragoza, 50009, Spain}
\author{Javier Campo}
\email[]{javier.campo@csic.es}
\affiliation{Aragon Nanoscience and Materials Institute 
  (CSIC -- University of Zaragoza) and
  Condensed Matter Physics Department, University of Zaragoza \\
  C/Pedro Cerbuna 12, Zaragoza, 50009, Spain}

\date{November 16, 2020}

\maketitle


In this supplemental material we provide details on the asymptotic solution of the
magnon spectral problem for the isolated chiral soliton in monoaxial helimagnets.
We also provide some details about the numerical calculations and show some numerical
results that complement those discussed in the paper.

\section{Asymptotic solution of the magnon spectral problem}

Much insight on the spectrum of $\Omega$ is obtained by analyzing the spectral equation
in the asymptotic regime, $z\to\pm\infty$. Although this is standard matter, it is
worthwhile to give some details of the computations that are relevant to the results
described in the paper.

\begin{figure}[t!]
\centering
\includegraphics[width=0.45\textwidth]{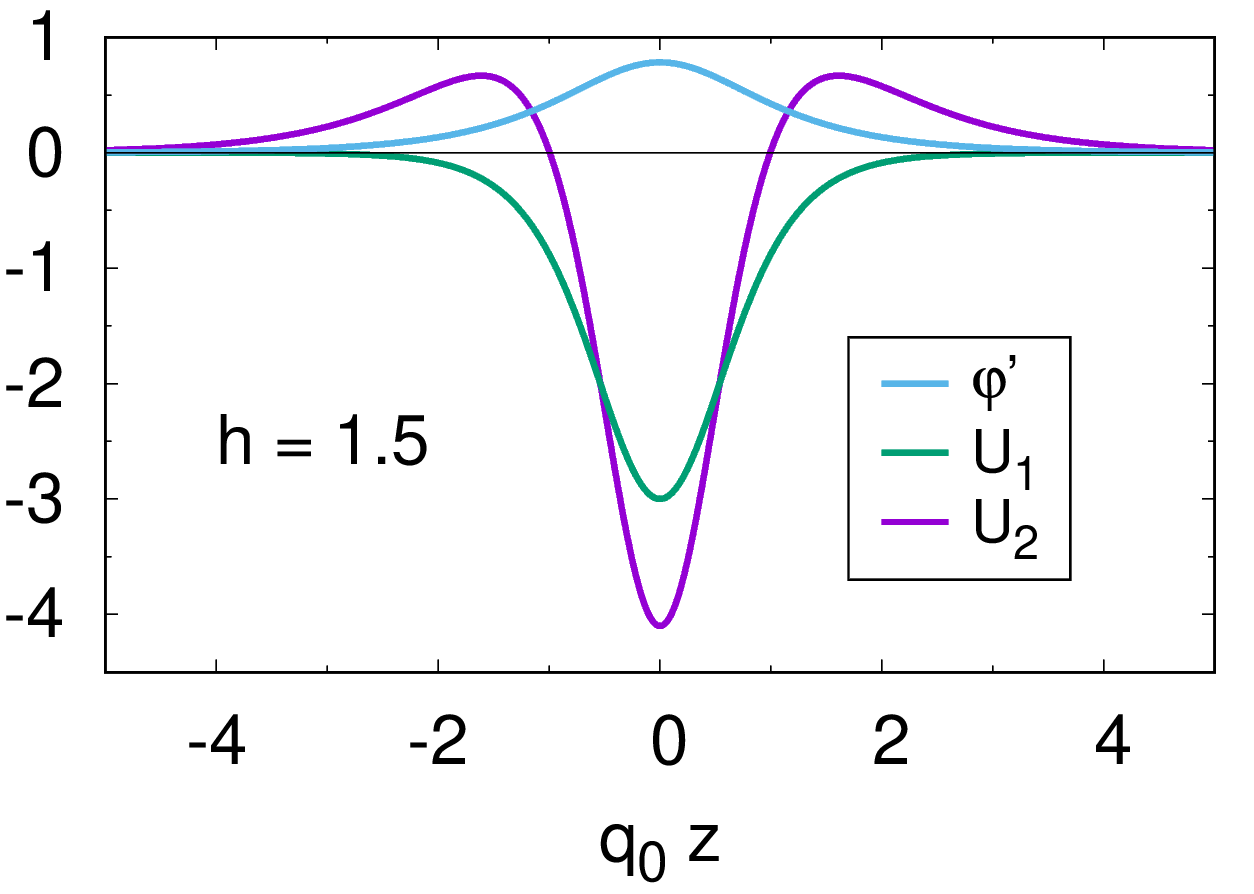}
\caption{Soliton profile, $\varphi^\prime$, and the potentials $U_1$ and $U_2$ for
  $\kappa=-5.0$ and $h=1.5$.
\label{fig:pots}}
\end{figure}

For the reader convenience, let us recall the form of $\Oik$ and $\Oiik$:
\begin{equation}
\Oik = \frac{\omega_0}{q_0^2}\big[-\partial_z^2 + U_1 + k_x^2 + q_0^2h\big], \quad
\Oiik = \frac{\omega_0}{q_0^2}\big[-\partial_z^2 + U_2 + k_x^2 + q_0^2(h-\kappa)\big]. 
\end{equation}
where \mbox{$U_1= -(1/2)\varphi^{\prime\,2}$} and
\mbox{$U_2 = -(3/2)\varphi^{\prime\,2}+2q_0\varphi^\prime$},
with \mbox{$\varphi^\prime(z)=2/[\Delta\cosh(z/\Delta)]$}.
Fig.~\ref{fig:pots} shows these functions for $\kappa=-5.0$ and $h=1.5$.

For $z\rightarrow\pm\infty$ the ``potentials'' $U_1$ and $U_2$ tends to zero exponentially
and the spectral equation $\Oiik\Oik\phi_{k_x}=\omega^2\phi_{k_x}$
asymptotically becomes
\begin{equation}
  \big[\partial_z^2-k_x^2-q_0^2(h-\kappa)\big]\big[\partial_z^2-k_x^2-q_0^2h\big]\phi_{k_x}
  = \frac{q_0^4\omega^2}{\omega_0^2}\phi_{k_x}. \label{eq:asymp}
\end{equation}
The solutions are exponential functions that can in general be written as $\exp(\iu k_zz)$,
for some $k_z$.
Equation (\ref{eq:asymp}) imposes a relation between $k_z^2$ and $\omega^2$, which can be
written as $\omega^2 = \omega_2\omega_1$, where
\begin{equation}
  \frac{\omega_1}{\omega_0} = \frac{k_z^2+k_x^2}{q_0^2}+h, \quad
  \frac{\omega_2}{\omega_0} = \frac{k_z^2+k_x^2}{q_0^2}+h-\kappa. \label{eq:omega}
\end{equation}
This relation can be inverted to give
\begin{equation}
  \frac{k_z^2}{q_0^2} = -\left(h+\frac{k_x^2}{q_0^2}-\frac{\kappa}{2}\right)
  \pm \left(\frac{\omega^2}{\omega_0^2}+\frac{\kappa^2}{4}\right)^{1/2}. \label{eq:kz}
\end{equation}
The right-hand side of the above equation is a real quantity.

Continuum states, unbounded in the $z$ direction, require $k_z$ real, that is $k_z^2\geq0$.
This condition requires to take the plus sign in equation (\ref{eq:kz}) and sets
a lower bound on $\omega$, written as $\omega>\omega_\mathrm{G}$, where
\begin{equation}
\omega_\mathrm{G}(k_x) = \omega_0\big[(k_x^2/q_0^2+h)(k_x^2/q_0^2+h-\kappa)\big]^{1/2}
\end{equation}
is the gap reported in equation (13) of the paper.
The continuum states are conveniently labeled by the wave number $k_z$, whose relation
with the eigenvalue $\omega^2$ is obtained from equation (\ref{eq:kz}):
\begin{equation}
  k_z = q_0\left[\left(\frac{\omega^2}{\omega_0^2}+\frac{\kappa^2}{4}\right)^{1/2}
    -\left(\frac{k_x^2}{q_0^2}+h-\frac{\kappa}{2}\right)\right]. \label{eq:kzcont}
\end{equation}
where $\omega\geq\omega_\mathrm{G}$ and we consider only $k_z\geq 0$.
Since the operator $\Oiik\Oik$ commutes with the
parity operator that implements the transformation $z\to -z$, its egenfunctions
are even and odd functions of $z$. The two degenerate values of the wave number,
$\pm k_z$ are combined to make the eigenfunctions with definite parity. Thus,
the continuum states are labeled by $k_z\geq 0$ and the parity, denoted by the
symbols $e$ (even) and $o$ (odd), so that we write
$\phi^{(e)}_{k_xk_z}(z)$ and $\phi^{(o)}_{k_xk_z}(z)$ for the eigenfunctions
of $\Oiik\Oik$.

Continuum states start at $k_z=0$, where $\omega=\omega_\mathrm{G}$, and fill the whole frequency
region above the gap.
For $k_x=0$ the gap is $\omega_\mathrm{G}(0)=\omega_{\mathrm{G}0}$, with
\begin{equation}
\omega_{\mathrm{G}0} = \omega_0[h(h-\kappa)]^{1/2}. \label{eq:oG0}
\end{equation}

Bound states in the $z$ direction require $k_z^2<0$ (imaginary $k_z$). There are two
possibilities: either the minus sign is taken in Eq.~(\ref{eq:kz}), in which case there is no
restriction in $\omega$, or the plus sign is taken and $\omega<\omega_\mathrm{G}$. In the latter
case the bound states are below the gap, while in the former bound states above the gap are
possible. The numerical results show that, for fixed $k_x$, there is a single bound state,
with even parity, located below the gap. At $k_x=0$ it is the zero mode associated to the
translation invariance of the soliton, and has $\omega=0$ and $k_z=\iu/\Delta$.
Thus the bound state branch is gapless.

\section{Details on numerical computations}

The spectral problems $\Oiik\Oik\phi_{k_x}=\omega^2\phi_{k_x}$
were solved numerically for a large discrete set of $k_x$,
on a box $-L\leq z \leq L$ with Dirichlet boundary conditions at $z=\pm L$,
that is, $\phi_{k_x}(\pm L)=0$.

The operators $\Oik$ and $\Oiik$ were discretized in the simplest way, with a symmetric
difference scheme for the second derivative, which guarantees hermiticity.
The spectrum of the discretized operator $\Oiik\Oik$ was obtained using the
linear algebra package ARPACK \cite{ARPACK}.
In practice, we found it more efficient to make use of the parity symmetry to restrict
the operators to the $(0,+L)$  interval, and obtain the even and odd spectrum separately,
using the boundary conditions appropriate for each case:
$\phi_{k_x}^{(e)}(-dz)=\phi_{k_x}^{(e)}(+dz)$ for the even eigenfunctions,
where $dz$ is the discretization step, and $\phi_{k_x}^{(o)}(0)=0$ for the odd eigenfunctions.
The computation were repeated for several values of $L$ and $dz$ to ensure that the results
show no noticeable volume or discretization effects.

The phase shifts are computed as follows: from the eigenvalue $\omega^2$ we compute the wave
number $k_z$ using Eq.~(\ref{eq:kzcont}). Then, the asymptotic condition given by
Eq.~(14) of the paper and the boundary condition at $z=+L$ gives the equation
$k_zL+\delta_i=2\pi n_i$, where $i=0,1$, and $n_i$ is the integer that makes
$\delta_i\in [-\pi,\pi]$.

\section{Some numerical results}

We show here some results that complement those described in the paper.

\begin{figure}
\centering
\includegraphics[width=0.45\textwidth]{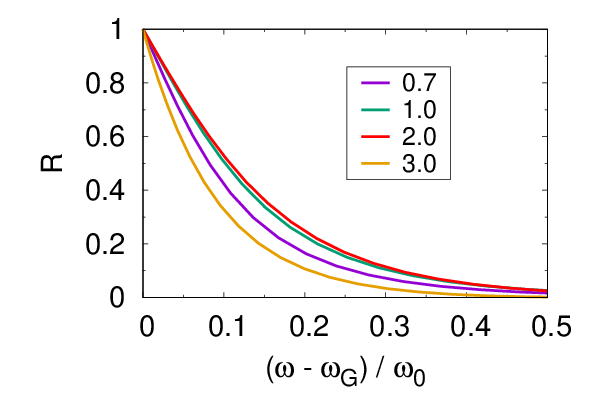}\*
\includegraphics[width=0.45\textwidth]{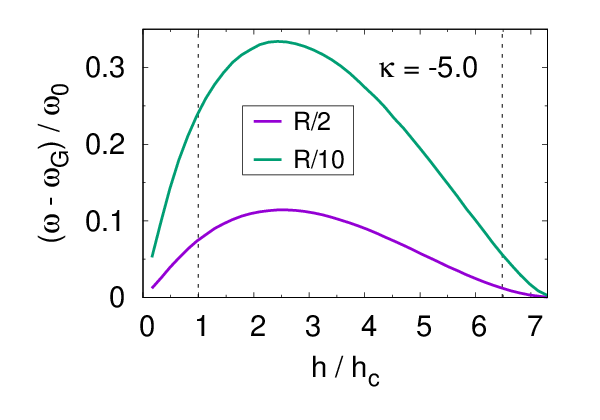}\*
\caption{Left: Reflection coefficient as a function of the frequency relative to the gap
  frequency, $\omega-\omega_\mathrm{G}$ for the values of $h$ displayed in the legend.
  Right: frequency, relative to the gap, at which the reflection coefficient decreases to 1/2
  (violet) and 1/10 (green), as a function of $h/h_c$.
\label{fig:refKz}}
\end{figure}

The reflection coefficient, given by $R=\sin^2(\delta_0-\delta_1)$, is displayed in
figure~\ref{fig:refKz} (left), for different values of the magnetic field.
It tends to zero as the frequency grows, as expected. The range
of frequencies, relative to the gap frequency, at which reflection is appreciable
depends non monotonically on the external magnetic field. That means there is a field
strength at which reflection is maximized, as illustrated in figure \ref{fig:refKz}
(right).

\begin{figure}[t!]
  \centering
  \includegraphics[width=0.485\textwidth]{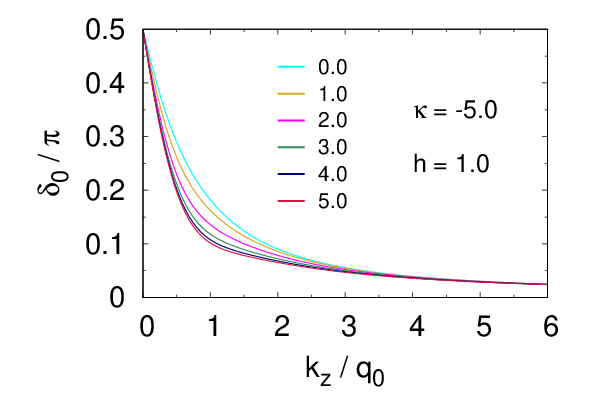}\*
  \includegraphics[width=0.485\textwidth]{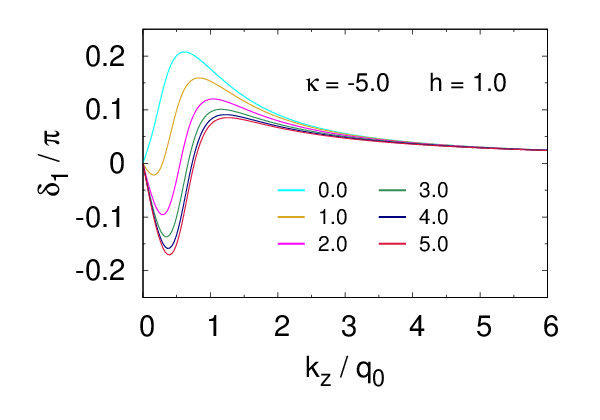}\*
  \caption{Phase shifts $\delta_0$ (left) and $\delta_1$ (right) as a function of
    $k_z$ for the values of $k_x/q_0$ displayed in the legend, with $\kappa=-5.0$ and $h=1.0$.
    \label{fig:pskx}}
\end{figure}

The dependence of the phase shifts on $k_x$ is illustrated in Figs.~\ref{fig:pskx}
for $\kappa=-5.0$ and $h=1.0$, were $\delta_0$ and $\delta_1$ are plotted as a function
of $k_z$ for several values of $k_x$. We notice that in all cases we have
$\delta_0(0)=\pi/2$, $\delta_1(0)=0$, and both $\delta_0$ and $\delta_1$ vanish as
$k_z\to\infty$. This means that in all cases the phase shifts are compatible with
Levinson theorem \cite{Galindo90I,Levinson49}.

\begin{figure}[t!]
  \centering
  \includegraphics[width=0.6\textwidth]{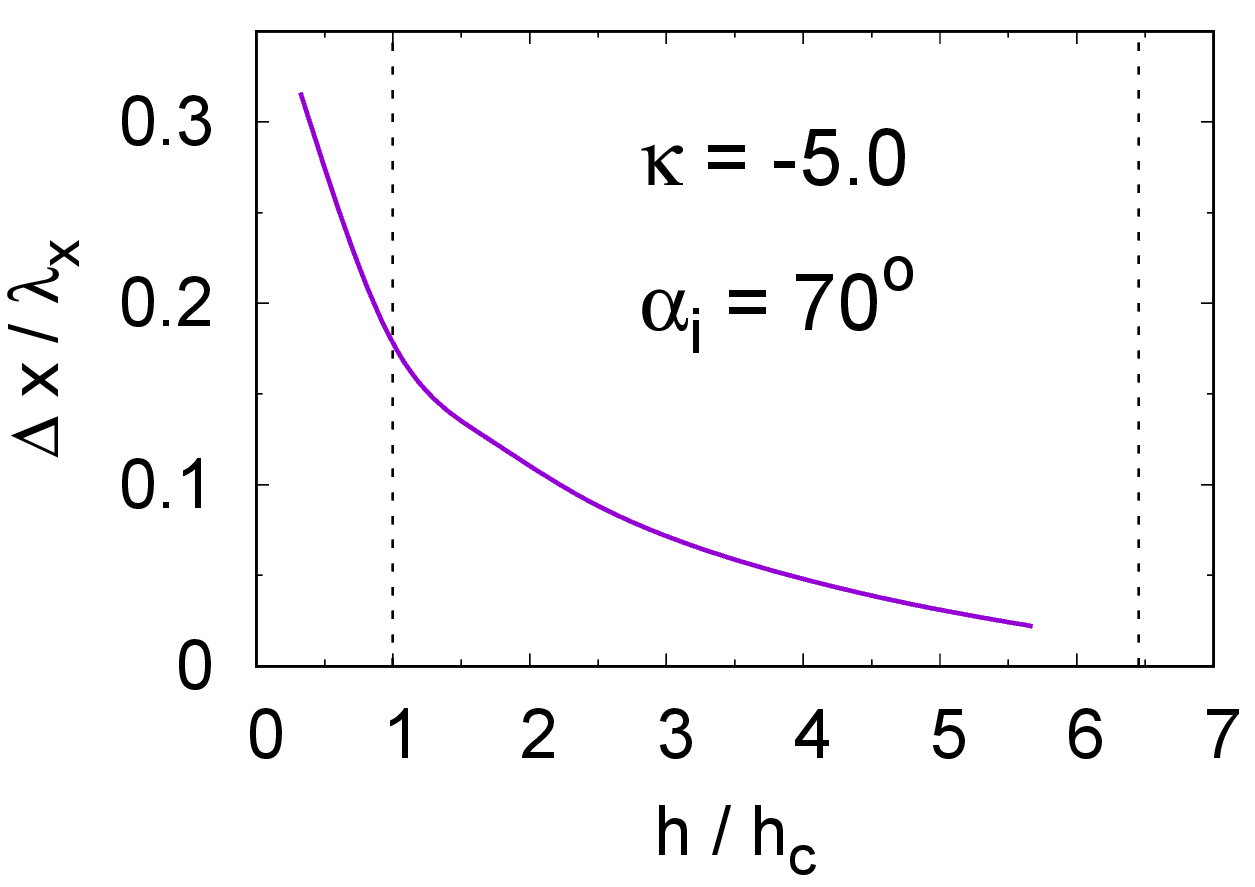}\*
  \caption{Maximum Goos-H\"anchen displacement for incidence angle
    $\alpha_i=70^\mathrm{o}$ vs. $h/h_c$, for $\kappa=-5.0$. The vertical dashed lines
    mark the critical field ($h=h_c$) and the destabilizing field.
    \label{fig:maxgh}}
\end{figure}

As stated in the paper, the Goos-H\"anchen displacement induced by the $k_x$ of the
phase sifts decreases with the applied field. Fig.~\ref{fig:maxgh} shows the
maximum displacement for an incidence angle of 70$^o$,
as a function of $h/h_c$ for $\kappa=-5.0$. The vertical dashed lines signal the position
of the critical field ($h/h_c=1$ and of the field strength at which the soliton
becomes unstable. The displacement vanishes at this destabilizing field.

\bibliography{references}